\documentclass[aps,pra,onecolumn,groupedaddress,tightenlines,nofootinbib]{revtex4}
\usepackage{amssymb}
\usepackage{color,graphicx}
\usepackage{lmodern}
\usepackage{amsmath}
\usepackage{amsbsy}
\usepackage{float}
\usepackage{bbm}
\usepackage{bm}
\usepackage{epsfig}
\usepackage{braket}
\usepackage{graphics}
\usepackage{subfigure}
\usepackage{appendix}
\usepackage{mathrsfs}

\newcommand\D{\operatorname{d}}
\newcommand\EXP{\operatorname{exp}}
\newcommand\ARG{\operatorname{arg}}
\newcommand\tr{\operatorname{Tr}}
\newcommand\Int{\operatorname{int}}
\newcommand\Ext{\operatorname{ext}}
\newcommand\NZ{\operatorname{NZ}}
\newcommand\TCL{\operatorname{TCL}}
\DeclareMathOperator{\sinc}{sinc}
\DeclareMathOperator{\disp}{disp}

\begin{document}
\title{Open Quantum System Approaches to Superconducting Qubits}
\author{Hamid Reza Naeij}
\email[]{hnaeij@mail.uni-paderborn.de}
\affiliation{Iranian Quantum Technologies Research Center (IQTEC), Tehran, Iran\\Department of Computer Science and Institute for Photonic Quantum Systems (PhoQS), Paderborn University, Germany}

\begin{abstract}

Superconducting qubits are currently a leading platform for developing a scalable quantum computer. However, random and uncontrollable noises from the environment during the design and measurement of qubits lead to limitations in qubit coherence time and gate fidelity, which is a major challenge in current state-of-art for superconducting quantum computing. To advance superconducting qubits technologies, it is essential to understand and mitigate environmentally-induced errors. This requires modeling superconducting qubits as open quantum systems coupled to their surroundings. The present study aims to provide useful open quantum system approaches to analyze and quantify the interaction between superconducting qubits and their environments. We provide an accessible introduction to open quantum system formalism for newcomers to the field. For experts, we discuss recently developed methods for analyzing superconducting qubit dynamics under realistic noises. We outline how these techniques provide quantitative insights into decoherence mechanisms and how they can guide design improvements to enhance qubit coherence time. This self-contained review of open quantum system approaches can be used to model, understand, and improve superconducting qubit performance in the presence of unavoidable environmental noises.

\end{abstract}

\maketitle 
\textbf{Keywords} decoherence, superconducting qubit, noise, circuit quantum electrodynamics 

 \tableofcontents

\section{Introduction}

Quantum computers are considered a next-generation technology that utilizes quantum properties such as superposition and entanglement to solve problems too complex for classical computers \cite{Feynman,Mermin,Nielsen}. Over the past twenty years, quantum computers have received significant attention across many fields including physics, chemistry, material science, communications, computer science, etc \cite{Bassman,Jager,Chertkov,Jerbi,Konopik,Naeij9}. 

The basic unit of quantum computing and quantum information is a two-state quantum system known as a quantum bit, {\it qubit}. Qubits can perform many calculations simultaneously, leading to a significant increase in computational speed. The quantum state of a qubit is a linear superposition of two basis states $\vert 0\rangle$ and $\vert 1\rangle$:

\begin{align}
\vert \psi \rangle= \alpha \vert 0\rangle+\beta \vert 1\rangle
\end{align}
where $\alpha$ and $\beta$ are complex numbers satisfying $\vert \alpha \vert ^2+\vert \beta \vert^2=1$. The relative quantum phase $\phi$ of the two basis states is $\phi=\ARG (\alpha^* \beta)$. A certain value of $\phi$ indicates phase coherence (or coherence) between the states $\vert 0\rangle$ and $\vert 1\rangle$ \cite{Nielsen}.

In recent years, scientists have made major advancements in developing quantum computers using different physical implementations of qubits, including cavity quantum electrodynamics \cite{Paternostro}, trapped ions \cite {Kielpinski}, NMR \cite{Jones}, ultracold atoms \cite{Gross}, quantum dots \cite{Zwanenburg} and nitrogen vacancies in diamonds \cite{Dutt}. To reach a practical and error-corrected quantum computer, increasing the number of qubits is a very important factor. To enable scalability, solid-state qubits are desirable. In this context, superconducting quantum systems are a promising solid-state approach, as they allow various options for qubit implementation to create quantum circuits \cite{Kwon}. 

Nowadays, superconducting qubits are one of the most important technologies used for realizing a universal quantum computer. However, the short coherence times of qubits due to various noises and the qubit interaction with the environment reduce the fidelity of quantum computing and pose significant challenges in superconducting qubit designs \cite{Kjaergaard,Krantz}, like other approaches in current state-of-art for quantum computing.

To build a practical quantum computer, the coherence time of qubits needs to be increased, which requires reducing environmental effects on the qubits. Understanding qubits dynamics and their interaction with the environment is essential for developing new computers and error correction techniques. However, precisely describing qubit-environment interaction is very challenging and requires noise spectroscopic experiments \cite{Burnett1,Lisenfeld}. Importantly, the coherence times vary significantly depending on the qubit type and materials.

The interaction between a quantum system and its environment leading to decoherence and the quantum-to-classical transition has been studied for decades in foundational and applied quantum mechanics \cite{Caldeira,Schlosshauer1,Breuer}. Decoherence poses a major challenge in implementing quantum information processing devices like quantum computers. Protecting quantum coherence is crucial for quantum computing, so examining models of qubit-environment interaction is very important. Consequently, many studies have focused on identifying decoherence sources, mechanisms and ways to mitigate decoherence effects and increase qubit coherence times in superconducting quantum computers, both theoretically and experimentally \cite{Cole,Muller,Burnett2,Eickbusch,Ficheux}. In this paper, we review useful approaches for analyzing qubit-environment interactions in superconducting qubits.

In the remainder of the paper, we first describe the physics of open quantum systems and decoherence theory. Then, by reviewing the superconducting qubits, we investigate the source of decoherence in these qubits and analyze the formalism of circuit quantum electrodynamics. Finally, we describe important open quantum system approaches to analyze the dynamics of superconducting qubits interacting with their environments.

\section{Decoherence}

A realistic quantum system is never completely isolated from its environment and interacts continuously with it. A schematic of an open quantum system is given in FIG. 1. The interaction between the system and the environment leads to the system being entangled with many environmental degrees of freedom. This affects the system dynamics and the results of measurements on the system. In particular, quantum coherence becomes effectively suppressed through a process called {\it Decoherence} that originates from the leakage of quantum information into the environment \cite{Schlosshauer1,Breuer}. Due to many uncontrollable degrees of freedom of the environment, the entanglement between the system and the environment is irreversible. 

Decoherence acts as a filter on the quantum states of a system. It selects certain states that can be easily prepared and maintained for a specific system while removing the superposition states. Thus, it can explain the quantum-to-classical transition and the measurement problem that are the heart of quantum mechanics \cite{Schlosshauer2, Schlosshauer3,Schlosshauer4}. 

\begin{figure}
\centering
\includegraphics[scale=0.3]{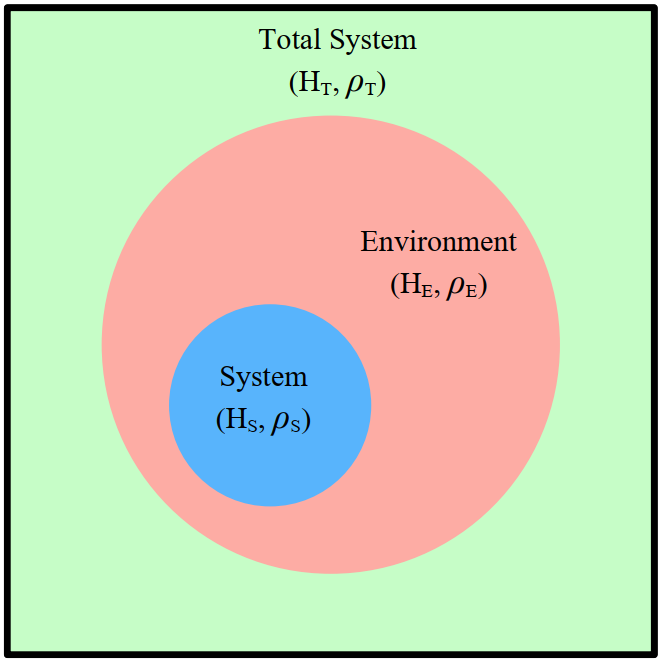}
\caption{A schematic of an open quantum system. A real quantum system is never completely separated from its environment. The interaction between the system and the environment leads to suppressing the quantum coherence of the system and affects the results of measurements on the system. In this figure, $H$ and $\rho$ denote the Hamiltonian and density matrix of the state, respectively.}
\end{figure}
 
Historically, the first evidence of the role of the environment in the description of quantum system behavior comes back to Heisenberg's writings \cite{Schlosshauer5}. In particular, the first paper on decoherence theory was published by H. Dieter Zeh in 1970 entitled "On the interpretation of measurement in quantum theory" \cite{Zeh1}. In this paper, Zeh stated that realistic quantum systems are never completely isolated and interact strongly with their environments. So, if we assume that the Schrödinger equation is universally valid, then these systems will usually be in states that are coupled with the environment in a quantum way. The term "Decoherence" emerged in the 1980s when Wojciech Zurek (a postdoc of John Wheeler’s at Caltech) and Erich Joos (a student of Zeh’s at Heidelberg) developed the formalism of decoherence \cite{Zurek1,Zurek2,Joos1}. Since then, various decoherence models and formalisms have been presented across the foundations of quantum mechanics and quantum technologies. It seems that all these efforts try to elucidate the multifaceted nature of decoherence. The literature in this field is vast \cite{Naeij1,Naeij2,Naeij3,Naeij4,Naeij5,Naeij6,Naeij7,Naeij8,Zurek3,Shu,Mondal,Youssefi,Cai,Soltanmanesh1,Soltanmanesh2}. 

The influence of the environment on the quantum system leads to disturbance of the system phase, {\it dephasing}, and energy exchange between the system and the environment, {\it dissipation} or {\it relaxation}. These processes feature in decoherence rate calculations. Moreover, experimental progress enabled decoherence observations in the matter-wave interferometry \cite{Hornberger}, superconducting systems \cite{Leggett} and cavity QED \cite{Raimond}. 

Note that decoherence is a purely quantum effect that should be distinguished from classical fluctuations. The important point is that decoherence is merely a consequence of applying the standard formalism of quantum mechanics to the interaction between a system and its environment, so it is not an external theory distinct from quantum mechanics \cite{Schlosshauer2}. However, for a comprehensive physical description of a quantum system in nature, it seems very necessary. 

In the following, we will explain the decoherence formalism specifics for superconducting qubits.

\section{Superconducting Qubits}

\subsection{Basic Definitions}

To date, various qubits have been designed for the development of quantum computers which differ in structure and performance \cite{Paternostro,Kielpinski,Jones,Gross,Zwanenburg,Dutt}. One of the most important and popular of them is superconducting qubits which are currently the leading technology for the physical realization of a scalable quantum computer. A superconducting qubit is the two lowest energy eigenstates of an artificial atom made of a superconducting circuit. A superconductor is a macroscopic quantum system that is defined as a substance that offers no resistance to the electric current when it becomes colder than a critical temperature \cite{Kwon}. Aluminum, copper oxide, magnesium diboride, yttrium barium, and niobium are well-known examples of superconductors. The size of superconducting circuits is macroscopic and quite different compared to other technologies used in quantum computers, such as quantum dots and trapped ions in which quantum information is encoded in microscopic systems such as electrons and photons. Macroscopicity of superconducting qubits means that they are not linked to the control of a single particle such as individual atoms, electrons or photons \cite{Yan,Barends,Oliver,Rasmussen}.

To review the historical background of superconducting qubits, we should go back to 1957 in which the Bardeen-Cooper-Schrieffer theory explained the superconductivity as a macroscopic effect that results from the condensation of Cooper pairs-opposite spin electrons-at low temperatures \cite{Bardeen}. Five years later, in 1962, the Josephson effect was discovered by Brian Josephson \cite{Josephson}. In the early 1980s, Leggett modeled the collective degrees of freedom of superconducting circuits \cite{Leggett1}. In 1985, Clarke, Devoret, and Martinis showed that a current-based Josephson junction can move from its zero-voltage state through a process called macroscopic quantum tunneling \cite{Martinis}. In 1998, the first Cooper Pairs Box (CPB) was designed and its ground state characterized \cite{Bouchiat}. In 1999, Nakamura, Pashkin and Tsai demonstrated the quantum coherent superposition with the first excited state \cite{Nakamura}. It was the first charge qubit, despite its short coherence time. In 2002, the CEA-Saclay Quantronics team presented the first practical version of the CPB, called the quantronium \cite{Vion}. In 2007, the modern version of the CPB circuit, the transmon, was designed at Yale University \cite{Koch}. Since then, various superconducting qubits with different structures, mechanisms, and coherence times have been presented such as cat-qubits, Unimon superconducting qubits, and Xmon tunable qubits \cite{Ezratty}.

\begin{figure}[H]
\centering
\includegraphics[scale=0.36]{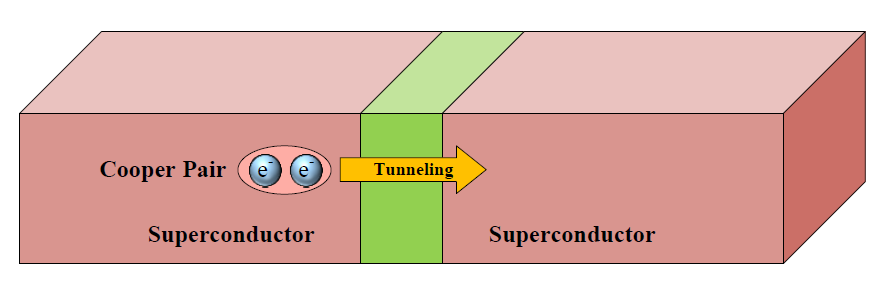}
\caption{A schematic of a Josephson junction. Two superconducting metals are separated by a thin insulator and Cooper pairs which tunnel between them.}
\end{figure}

To design superconducting qubits, in addition to the superconductivity, a confinement potential is needed to have discrete energy eigenstates. Furthermore, to selectively control the two lowest-energy states, the potential has to be anharmonic, so that there is a distinct energy separation between the states. To generate the anharmonicity, a Josephson junction is used in superconducting qubits. A Josephson junction is a pair of superconductors that are weakly coupled. As is given in FIG. 2, a Josephson junction is a nanometer-thin insulating barrier between two superconducting metals that creates a tunnel junction. This produces a quantum electrical component, the superconducting phase difference between its electrodes, that is conjugated to the number of Cooper pairs passing through the junction. In simple terms, a Josephson junction behaves as a non-linear and non-dissipative inductance in a superconducting circuit and its value changes based on the phase and therefore current flowing through it. At very cold temperatures, lower than the critical temperature for superconductivity, Josephson junctions in an electrical circuit act like an artificial atom. They have different quantum levels that can be controlled by a gate or a flux, and they contain around 100 billion electron Cooper pairs. 

As is shown in FIG. 3, the non-dissipative components of superconducting qubits are capacitors, inductors, and the Josephson junction. Capacitors store energy in the electric field, while inductors store energy in the magnetic field \cite{Kwon, Ezratty}.

\begin{figure}
\centering
\includegraphics[scale=0.37]{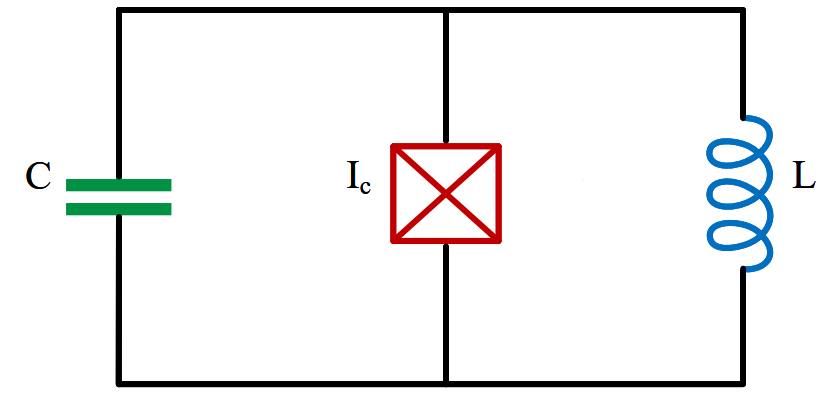}
\caption{A general schematic of a superconducting qubit composed of an inductor (L), a capacitor (C) and a Josephson junction with critical current (I$_c$). Source \cite{Kwon}.}
\end{figure}

The most important distinction between different types of superconducting qubits arises from the qubit state encoding method. Three important categories of superconducting qubits are \cite{Kwon,Rasmussen,Ezratty}:

\subparagraph{Charge qubits} in which the ground states of the qubit are eigenstates of charge. It means that they are only dependent on the number of excess Cooper pairs in a disconnected superconducting island. Single CPB and transmon are well-known examples. The transmon significantly improved upon earlier charge qubits by operating in a regime with a much larger ratio of Josephson energy to charging energy. This design choice exponentially suppresses the qubit's sensitivity to charge noise, a major source of dephasing in earlier CPBs, leading to substantially longer coherence times, albeit at the cost of reduced anharmonicity. Moreover, it shows well-developed control and readout schemes using microwave signals.

\subparagraph{Flux qubits} consist of a superconducting loop interrupted by one or more Josephson junctions. Their quantum states are typically superpositions of states corresponding to persistent superconducting currents flowing in opposite directions around the loop, creating a magnetic flux. Flux qubits normally exhibit large anharmonicity, comparable to or greater than early CPB charge qubits. Compared to transmon qubits, which sacrifice some anharmonicity for enhanced charge noise insensitivity, flux qubits can offer a different balance of these properties and often achieve high coherence. C-shunted flux qubits and fluxonium qubits are examples. Unlike the transmon, the fluxonium typically incorporates a junction array or a large superinductor, shunting a small Josephson junction. This design creates a different energy spectrum, characterized by a small plasma frequency and strong anharmonicity. Fluxonium qubits can exhibit excellent coherence times, particularly robust against flux noise at specific operating points, and offer a rich parameter space for engineering qubit properties. However, it requires complex fabrication (superinductor is large and made of many junctions) and more complicated control/readout than transmon due to a more complex energy spectrum.

\subparagraph{Phase qubits} which have bigger Josephson junctions compared to charge qubits. Their states relate to two levels of current energy in a Josephson junction which two levels are defined by quantum oscillations of the phase difference between the two sides of the junction. In phase qubits, the different states are usually determined by the direction of a flowing current in the Josephson loop. Phase qubits were among the early successful demonstrations of coherent quantum phenomena in macroscopic superconducting circuits. Moreover, experiments with phase qubits in the early 2000s (e.g., by John Martinis' group) were pivotal in demonstrating single-qubit operations (initialization, Rabi oscillations, readout) and contributed significantly to establishing superconducting circuits as a viable platform for quantum computing \cite{Devoret}.

These important qubits can be seen in FIG. 4. For more details about types of superconducting qubits, please see \cite{Kwon,Rasmussen,Ezratty}.

\begin{figure*}[h]
\centering
\includegraphics[scale=0.37]{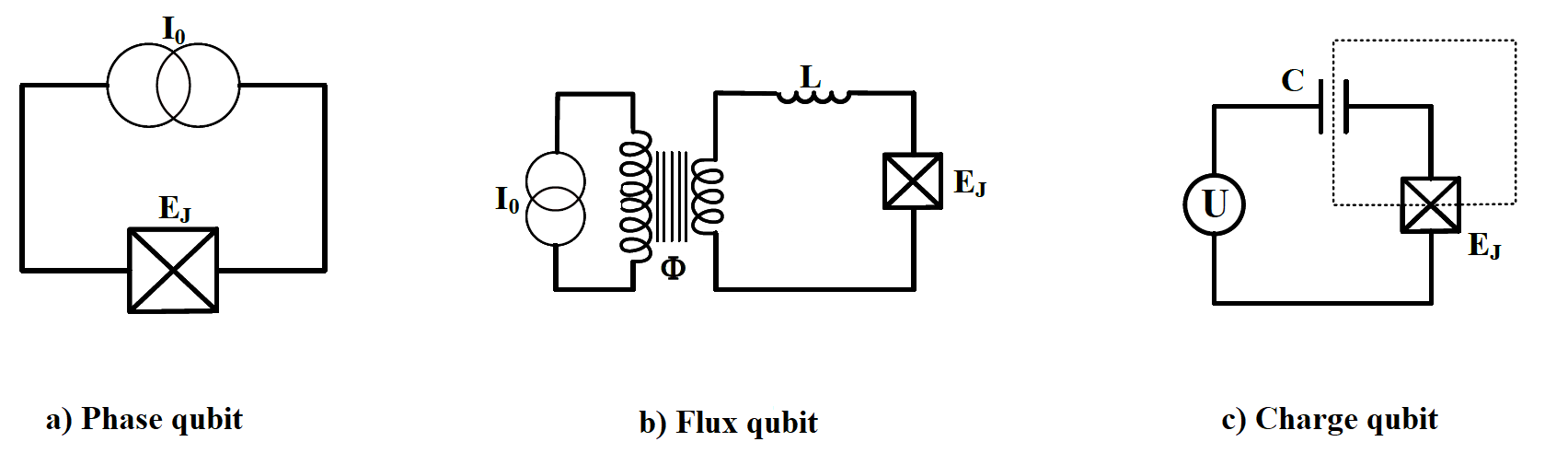}
\caption{The different types of superconducting qubits in which $E_J$, $U$, $\Phi$ and $I_0$ are the Josephson energy, voltage source, flux and current, respectively. Source \cite{Devoret}.}
\end{figure*}

\subsection{Hamiltonian}

In this section, we define a foundational Hamiltonian form, given in FIG. 3, that describes many common superconducting qubits based on a single Josephson junction with critical current (I$_c$) along with linear inductive (L) and capacitive (C) elements. Note that this Hamiltonian serves as a starting point for understanding various qubit types such as charge, flux, and transmon qubits, though more complex or specialized designs might necessitate extended or different Hamiltonians \cite{Kwon,Vool}.

To analyze the physics of the Josephson junction, we introduce two parameters: the number imbalance of electrons $N$ and the relative phase $\phi$ between the two superconductors. These parameters are related to each other by Josephson's equation of motion \cite{Leggett3}

\begin{subequations}
\begin{align}
\frac{dN(t)}{dt}&=\frac{2E_J}{\hbar}\sin \phi(t) \\
\frac{d\phi(t)}{dt}&=-\frac{2e}{\hbar}V(t),
\end{align}
\end{subequations}
where $E_J$ is the Josephson energy which quantifies the ability of Cooper pairs to tunnel through the junction. Moreover,  $\hbar$, $e$ and $V$ are the Planck constant, the magnitude of the charge carried by a single electron and the voltage difference across the junction, respectively. 

The first formula in Eq. (2) can be written as 

\begin{align}
I_s(t)=I_c \sin \phi(t),
\end{align}
where $I_s$ is a zero-voltage supercurrent flow in the Josephson junction and $I_c=2eE_J / \hbar$ is the critical current of the junction, which means the maximum current that can pass through the junction.

We can easily quantize the superconducting circuit by considering $N$ and $\phi$ as the operators $\hat N$ and $\hat \phi$. It is importyant to note $\hat N$ and $\hat \phi$ are conjugated as $\hat{N}=-i \partial / \partial \phi$ which satisfy the realation $e^{i\hat{\phi}}\hat{N}e^{-i\hat{\phi}}=\hat{N}-1$. From now on, throughout the paper, we ignore the operator sign for simplicity.

The general form of the superconducting circuit Hamiltonian can be defined as \cite{Kwon}

\begin{align}
H=4E_C (N-N_{\Ext})^2+\frac{1}{2} E_L \phi^2-E_J\cos (\phi-\phi_{\Ext}),
\end{align}
where the capacitive energy is $E_C=e^2/ 2C$ which represents the energy cost to charge a capacitor with a single electron and the factor of 4 denotes the Cooper pairing. $E_L=(\Phi_0/2\pi)^2/L$ is the inductive energy showing the energy cost to charge an inductor with a single flux quantum $\Phi_0$.  Moreover, $E_J$ shows the energy stored in the junction and can be obtained by integrating the electrical work $\int I_sV \D t$. $N_{\Ext}$ is the charge offset due to the external voltage bias and $\phi_{\Ext}=2\pi \Phi/ \Phi_0$ is the phase offset due to the external flux bias $\Phi$.  We assume that magnetic flux penetrates the loop through the junction, not through the inductor. Therefore, we can add the phase offset in $H$. As is clear in Eq. (4), the properties of a superconducting qubit can be engineered by three important circuit parameters: $E_J$, $E_C$, and $E_L$ \citep{Kwon,Rasmussen,Ezratty}.

\subsection{Decoherence in Superconducting Qubits}

As we discussed in section II, to fully describe superconducting qubits, we should consider their coupling with the surrounding environment. In quantum computing, describing qubits in complete isolation conditions is impractical, because we have no way to control or observe them \cite{Blais}.

Superconducting qubits are macroscopic states such as Schrödinger’s cat states, so they are very sensitive to environmental effects and noises. The unavoidable coupling between the qubits and the environment causes decoherence. Due to the unique characteristics of superconducting qubits, precise engineering is necessary to increase decoherence time and reduce noise effects. This issue is one of the most important challenges of designing quantum computers based on superconductors. Low decoherence time causes errors in quantum computing, so it is necessary that the coherence time of a qubit should be large enough to enable error correction. After decoherence time, the qubit stops working and doesn't work correctly anymore. However, the weak interaction between the qubit and its environment is necessary in the readout step of the qubits \cite{Catelani}.

\subsubsection{Formalism of decoherence}

The formalism of decoherence in superconducting qubits is similar to other quantum systems and includes two main steps: {\it relaxation} and {\it dephasing} \cite{Dong,Feng}. These mechanisms lead to the exponential vanishing of off-diagonal elements of the density matrix $\rho$ of the qubit which shows the overlap between two states of a qubit. 

To describe these processes, two experimentally measurable time scales are introduced: $T_1$ and $T_2$. The physical process responsible for $T_1$ is relaxation (or qubit thermalization) which occurs through incoherent energy exchange between the qubit and its environment. This interaction is often modeled as fluctuations in the Hamiltonian of qubit. In the Hamiltonian, Eq. (4), there are physical quantities that mediate the interaction between the qubit and the environment, such as external charge and magnetic flux bias. In $T_2$, in addition to the relaxation, the process called dephasing also contributes. Dephasing means the loss of phase coherence of the qubit and is caused by the temporal fluctuation in qubit transition energy, the energy of a quantum transition between two energy levels \cite{Kwon,Ficheux}. 

Relaxation and dephasing times can be formulated as \cite{Kwon,Krantz}

\begin{subequations}
\begin{align}
&\Gamma_1=\frac{1}{T_1} \\
&\Gamma_2=\frac{1}{T_2}=\Gamma_{\phi}+\frac{\Gamma_1}{2}
\end{align}
\end{subequations}
in which $\Gamma_1$ is the population decay rate of excited state and $\Gamma_{\phi}$ is the dephasing rate.

In the relaxation process, spontaneous emission and absorption are two main mechanisms. A qubit in state $\vert 1 \rangle$ emits energy to the environment and relaxes to state $\vert 0 \rangle$ and a qubit in state $\vert 0 \rangle$ absorbs energy from the environment and excites to state $\vert 1 \rangle$. However, if the qubit frequency is much larger than the temperature of the environment, i.e. $k_BT\ll \hbar \omega_q$, we can ignore the contribution of the absorption process in the calculation of $\Gamma_1$. Moreover, the environment should be able to dissipate electromagnetic energy near the frequency of qubit. Therefore, the relaxation is dominated by the noise whose frequency is near $\omega_q$ that demonstrates relaxation is a resonant phenomenon. On the other hand, in dephasing, the time scale of the fluctuation in transition frequency should be much slower than the qubit transition and the same magnitude as the measurement time scale. So, $\Gamma_{\phi}$ is mainly calculated by low-frequency noise. Consequently, the Ohmic noise and $1/f$ noise are responsible for $\Gamma_1$ and $\Gamma_\phi$, respectively \cite{Yan,Astafiev, Paladino,Yoshihara,Bylander}. Moreover, in contrast to the relaxation process, the energy of a qubit is conserved during dephasing. It means pure dephasing is elastic and reversible. Therefore, we can restore phase coherence and information that has leaked from the system into the environment by unitary operations such as applying pulses that can reverse the direction of time evolution \cite{Kwon, Krantz}. In FIG. 5, the details of the pathways to decoherence for superconducting qubits based on the Bloch sphere are given \cite{Krantz}.

\begin{figure*}
\centering
\includegraphics[scale=0.096]{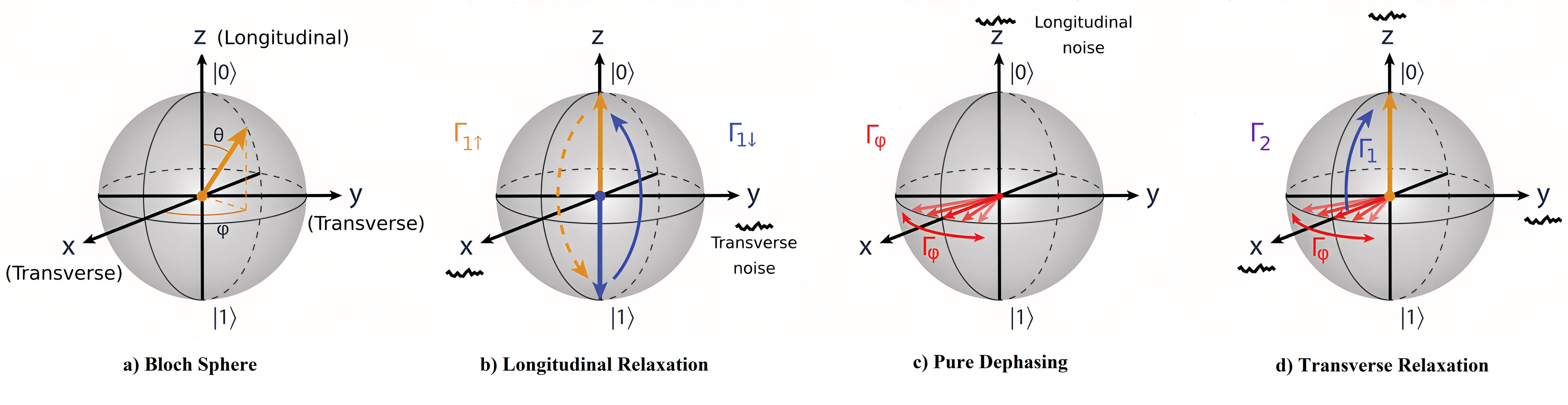}
\caption{A schematic of pathways to decoherence in a superconducting qubit. (a) Bloch sphere representation of a qubit $\vert \psi \rangle= \alpha \vert 0\rangle+\beta \vert 1\rangle$. The quantization axis of the qubit $z$-axis is longitudinal in the qubit frame and the $xy$ plane is transverse. (b) The coupling of transverse noise to the qubit in the $xy$ plane causes the energy exchange between the qubit and its environment called Longitudinal relaxation. Its result is $\vert 0 \rangle \rightarrow \vert 1 \rangle$. A qubit in state $\vert 1 \rangle$ emits energy to its environment and relaxes to $\vert 0 \rangle$ with a rate $\Gamma_{1\downarrow}$. Furthermore, when a qubit is in the state $\vert 0 \rangle$, it absorbs energy from its environment and excites to the state $\vert 1 \rangle$ at a certain rate, which is denoted as $\Gamma_{1\uparrow}$. If the qubit frequency is much larger than the temperature of the environment $k_BT\ll \hbar \omega_q$, we ignore the contribution of the absorption process in the calculation of $\Gamma_1$. So we have $\Gamma_1 \approx \Gamma_{1\downarrow}$. (c) The longitudinal noise along the $z$-axis causes pure dephasing in the transverse plane. It means that the azimuthal phase of the qubit will be depolarized with a rate $\Gamma_{\phi}$. (d) Transverse relaxation is a combination of energy relaxation and pure dephasing leading to loss of coherence at the decoherence rate $\Gamma_2$. Moreover, the excited state of the qubit may relax to the ground state. It is a phase-breaking process that leads to the loss of the orientation of the Bloch vector in the $xy$ plane. Source \cite{Krantz}.}
\end{figure*}

\subsubsection{Source of decoherence}

The source of decoherence in superconducting qubits is the stochastic noises resulting from the interaction between the qubit and the environment. The noise leads to unwanted terms in the Hamiltonian of the qubits, and consequently unwanted operations in the computing process. Therefore, the analysis of the dynamics and performance of the qubits would be complicated. 

There are many types of noise in superconducting qubits. However, we briefly discuss some of the most common types of them.

\subparagraph{Charge noise} is a common noise in solid-state devices. This noise originates fluctuating charges in the structural disorder or charge traps in the dielectrics and tunnel barrier of the junction. It can be modeled as an ensemble of fluctuating two-level systems or as bulk dielectric loss \cite{Wang,Dial}.  To mitigate the effects of charge noise, we can use the etching and passivation to remove amorphous layers, crystalline dielectrics and ordered Josephson junction barriers \cite{Siddiqi}.

\subparagraph{Photon number fluctuations} is the major decoherence source in the circuit quantum electrodynamics \cite{Schuster1}. Photon number fluctuations of microwave fields in the cavity affect the superconducting qubits in the dispersive limit that leads to pure dephasing ($T_2$) \cite{Krantz}.

\subparagraph{Quasiparticles and phonons} which are generated from incident ionizing radiation break Cooper pairs. Quasiparticles such as unpaired electrons tunnel through the junction leading to the relaxation ($T_1$) and pure dephasing ($T_2$). The effect of quasiparticle noise on the superconducting qubits depends on the type of qubit and the junction \cite{Catelani2,Gustavsson}. To mitigate the effects of this noise, we can use normal metal and low-gap superconducting traps to remove quasiparticles from the qubit area and acoustic absorbers to suppress phonon propagation \cite{Siddiqi}.

\subparagraph{$1/f$ magnetic flux noise} is the main decoherence source of superconducting qubits resulting from low-frequency noise. The spectral density of $1/f$ noise is inversely proportional to the frequency. This noise arises from fluctuating magnetic impurities, spins or clusters of spins on metallic surfaces, electrons trapped in disorder potentials, or paramagnetic films that condense at low temperatures \cite{Clarke}. To reduce the effects of $1/f$ noise, we can use the capping layers, suitable surface passivation and improvements in the sample vacuum environment \cite{Siddiqi,Kumar}.

Note that various types of superconducting qubits have different sensitivities to noise. For example, the coherence times of the phase and flux qubits are affected by flux noise while the transmon qubits are chosen to reduce the charge noise effect compared to the Cooper pair box \cite{Koch,Yoshihara,Bialczak}. Moreover, to mitigate the noise effects, several noise-resilient qubits are designed in which the robustness of the qubits can be improved by adjusting the superconducting circuit parameters \cite{Kwon}.

\section{Circuit Quantum Electrodynamics
(CQED)}

So far, we have explained the basic conceptual elements of superconducting qubits. To perform quantum computing with superconducting qubits, it is necessary to couple the qubits with other systems so that we can control and read the state of the qubits \cite{Kwon}. In this regard, a version of cavity quantum electrodynamics (QED) that is called circuit quantum electrodynamics (CQED) is formulated to analyze the physics of the interaction of superconducting circuits that act as qubits or artificial atoms with quantized electromagnetic fields in microwave resonators \cite{Blais}. QED is a well-known approach to study the interaction of atoms and the quantized electromagnetic modes inside a cavity and has attracted much attention in the study of open quantum systems, decoherence theory, and quantum information processing \cite{Raimond,Mabuchi,Hood,Blais2}.

CQED emerged as a highly influential research field approximately two decades ago, rapidly becoming a cornerstone for superconducting quantum computing \cite{Blais2,Wallraff,Girvin}. CQED uses theoretical and experimental methods to manipulate quantum states for technological applications. Currently, it plays an essential role in quantum information processing based on superconducting qubits \cite{Blais}. Moreover, CQED is now a leading strategy in single photon detectors \cite{Besse, Kono,Lescanne}, development of quantum-limited amplifiers \cite{Clerk1,Roy}, hybrid quantum systems \cite{Clerk2} and semiconducting quantum dots \cite{Burkard}.

A basic model to analyze the interaction between matter and light is a two-level atom interacting with a cavity through a dipole coupling. The fully quantum-mechanical version of this model is the Jaynes-Cummings model to describe the coherent behavior of the coupled system \cite{Knight,Walls}. Besides atoms and photons, there are many systems that can be represented by Jaynes-Cummings Hamiltonian, such as circuits \cite{Wallraff}, quantum dots \cite{Reithmaier,Yoshie} and nano-mechanical systems \cite{Irish}.

In CQED, the Jaynes–Cummings Hamiltonian analyzes how a superconducting qubit exchanges photons with its surrounding cavity. The atom-cavity interaction can be implemented by the qubit–resonator coupling and the circuit Hamiltonians for a qubit and a resonator are mapped to spin-$1/2$ fermionic and bosonic Hamiltonians, respectively. 

The total Hamiltonian $H$ of the qubit and the cavity in CQED formalism can be written as \cite{Kwon,Ezratty}

\begin{align}
H=\hbar \omega_{\rm c} (a^{\dagger} a+\frac{1}{2})+\frac{\hbar \omega_{\rm q}}{2}\sigma_z +\hbar g (a^{\dagger}\sigma^-+\sigma^+ a)+H_{\rm c\varepsilon} +H_{\rm q\varepsilon}.
\end{align}
The first term represents cavity harmonic oscillator energy that $\omega_{\rm c}$ is the angular oscillation frequency of the relevant cavity mode. The second term shows the qubit energy in which the qubit frequency $\omega_{\rm q}$ is the energy difference between the ground state $\vert g \rangle$ and the excited state $\vert e \rangle$ of the qubit. The third term denotes the coupling between qubit and cavity with the strength of interaction $g$ in which the rotating wave approximation (RWA) is applied \cite{Knight,Walls}. This term shows the coherent exchange of excitation between the qubit and cavity and describes a dipole interaction where an atom can absorb ($\sigma^+ a$) and emit ($a^{\dagger} \sigma^-$) a photon from/to the field at rate $g$. The two last terms show the coupling of cavity and qubit with the environment, respectively, that causes relaxation and dephasing of the qubit. Moreover, $a$ and $a^{\dagger}$ are annihilation and creation operators of the resonator photons, respectively and $\sigma_z=\vert e \rangle \langle e \vert- \vert g \rangle \langle g \vert$ is the atomic inversion operator and the operators $\sigma^-=\vert g \rangle \langle e \vert $ and $\sigma^+= \vert e \rangle \langle g \vert $ are the raising and lowering operators of the qubit. The detuning of a qubit from the cavity can be defined as $\omega_{\rm cq}=\omega_{\rm q}-\omega_{\rm c}$.

In Jaynes-Cummings Hamiltonian, there are many concepts such as the Jaynes-Cummings spectrum, the energy levels of the qubits (dressed states), the resonant regime of qubit-cavity, and enabling readout of qubits with resonator (dispersive regime) \cite{Xiu}. Moreover, some parameters characterize a superconducting qubit’s properties such as $Q$ factor, the ratio between the energy stored in an oscillator and the energy dissipated per oscillation cycle times $2\pi$. It represents the stability of the qubit and determines its relaxation time $T_1$. The higher $Q$ factor, the longer $T_1$ will be, although this can be detrimental to noise sensitivity \cite{Burnett2, Ezratty}.

Note that in the resonant regime, the qubit and the photon can exchange energy, which causes them to lose their individuality and form a strongly coupled system. However, in some cases, it is preferred that the interaction be dispersive means that no actual photons are absorbed by the qubit and the effect of the dipole interaction can be modeled using perturbation theory. In this regime, the qubit is far detuned from the cavity \cite{Schuster}.

In CQED, the Jaynes-Cummings model predicts results that are in qualitative agreement with the experiments, in particular only the first two levels of the qubit are important. However, the quantitative agreement between experimental results and theoretical predictions is obtained only when accounting for higher qubit energy levels and the multimode nature of the field \cite{Blais}. A schematic of CQED is given in FIG. 6.

\begin{figure*}
\centering
\includegraphics[scale=0.09]{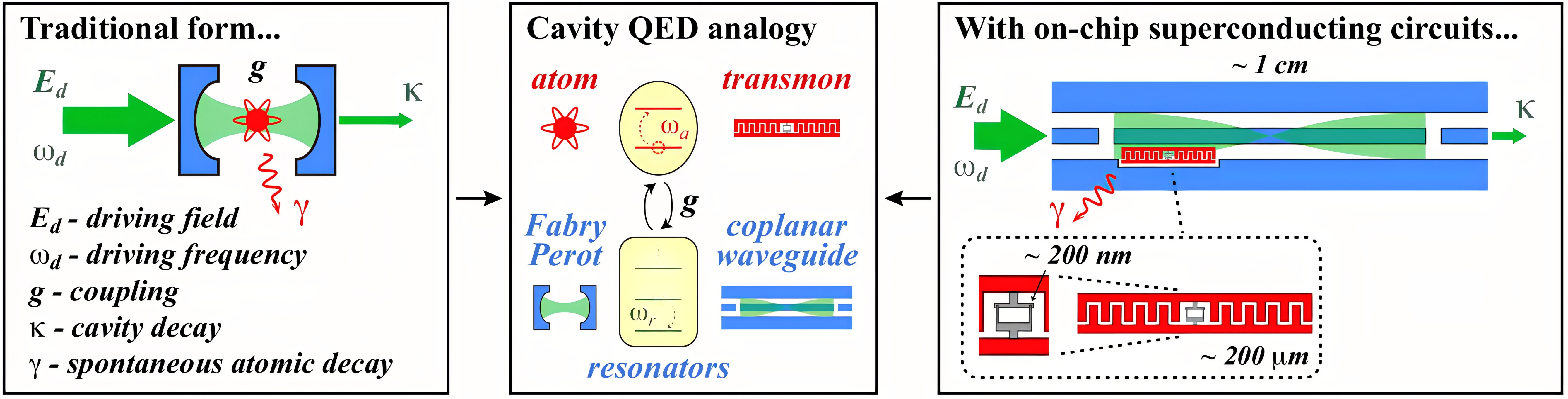}
\caption{A schematic of CQED in superconducting qubits based on cavity QED to describe the interaction between qubits and resonator. Source \cite{Langford}.}
\end{figure*}

\section{Approaches to Decoherence in Superconducting Qubits}

The coupling of superconducting qubits and the environment is vital for coherent control and measurement in CQED, but it always leads to decoherence. Knowledge about how qubits evolve in different operations is very important to implement functions required for quantum computing based on superconducting qubits. So far, various approaches have been proposed to analyze the dynamics of superconducting qubits interacting with the environment. In this section, we analyze open quantum system approaches to model qubit-environment interaction in CQED and other superconducting qubit systems.

\subsection{Lindblad Master Equation}

One of the most important approaches to modeling decoherence in the theory of open quantum systems is the master equation \cite{Schlosshauer1,Breuer}. In this regard, to analyze the effect of noise on the dynamics of superconducting qubits, that is an electrical circuit, we use a type of master equation called Lindblad master equation. In the following, we formulate the formalism of the Lindblad master equation.

\subsubsection{Non-unitary evolution}

As is known, the dynamics of a closed quantum system is described by the Schrödinger equation 

\begin{align}
i\hbar\frac{\D}{\D t}\left|\psi\right> = H \left|\psi\right>,
\end{align}
where $\left|\psi\right>$ and $H$ are the state vector and the Hamiltonian of the system, respectively. This relation is the Schrödinger equation in matrix form. This matrix equation represents the unitary evolution of a pure quantum system as

\begin{align}
\left|\psi (t)\right>= U \left|\psi (0)\right>,
\end{align}
where $U=\exp (-iHt/\hbar)$.

The interaction between the quantum system and its environment causes stochastic transitions between energy levels and uncertainty in the phase between states of the system. In other words, unlike the evolution of the state vector in a pure quantum system which is deterministic, the dynamics of open quantum systems is stochastic. 

To describe the state of an open quantum system as a mixed state, we use the density matrix formalism. A density matrix $\rho$ is a Hermitian operator describes a probability distribution of quantum states $\left|\psi_n\right>$  in a matrix representation that defined as $\rho = \sum_n p_n \left|\psi_n\right>\left<\psi_n\right|$, where $p_n$ is the probability that the system is in $\left|\psi_n\right>$. The normalization condition of the density matrix is defined as $\tr \rho=1$ where $\tr$ denotes Trace operation. Moreover, for a pure quantum state, we have $\tr \rho^2=1$ and for a mixed state $\tr \rho^2< 1$. In general, the time evolution of a density matrix $\rho$ is the basis of the master equations in the theory of open quantum systems \cite{Schlosshauer1,Breuer}.

\subsubsection{Derivation of the Lindblad master equation}

To analyze the dynamics of the qubit interacting with its environment (such as resonator), we consider the total quantum system (qubit+environment) to be closed and its evolution is governed by the Liouville-von Neumann equation \cite{Schlosshauer1,Breuer}

\begin{align}
\frac{\D \rho}{\D t} = -\frac{i}{\hbar}[H, \rho (t)],
\end{align}
where is the density matrix version of Schrödinger equation in Eq. (7) in which $\rho$ and $H$ denote the density matrix and Hamiltonian of the total system, respectively. The total Hamiltonian is written as

\begin{align}
H = H_{\rm q} + H_{\varepsilon} + H_{\Int},
\end{align}
in which $H_{\rm q}$, $H_{\varepsilon}$ and $H_{\Int}$ are the Hamiltonians of the qubit, the environment and the interaction between the qubit and its environment, respectively. Moreover, the Hilbert space of the total system can be defined as

\begin{equation}
\mathscr{H}=\mathscr{H}_{\rm q} \otimes \mathscr{H}_{\rm \varepsilon},
\end{equation}

To formulate the equation of motion of a qubit interacting with its environment, we use three important assumptions \cite{Schlosshauer1,Breuer}:

\subparagraph {i. Separability} At $t=0$, there are no correlations between the system and its environment. Therefore, the total density matrix can be defined as a tensor product $\rho(0) = \rho_{\rm q} (0) \otimes \rho_{\rm \varepsilon}(0)$.

\subparagraph {ii. Born approximation}The interaction between the qubit and the environment is sufficiently weak. Moreover, the environment is much larger than the system, so changes in the density operator of the environment are negligible and the system and the environment remain in an approximate product state at all times, i.e.,

\begin{align}
 \rho(t) \approx \rho_{\rm q}(t)\otimes\rho_{\rm \varepsilon},
\end{align}
where $\rho_{\rm \varepsilon}$ approximately constant at all times.

\subparagraph {iii. Markov approximation}Memory effects of the environment are negligible. In other words, the time scale of decay of environmental correlation functions is much shorter than the time scale of the qubit dynamics $\tau_{\rm q} \gg \tau_{\rm \varepsilon}$.

To obtain the dynamics of the qubit, we calculate the reduced density matrix of the qubit by performing a partial trace over the environmental degrees of freedom ($\rho_{\rm q}=\tr_\varepsilon \rho $) in Eq. (9). In this way, a master equation for the motion of the qubit is obtained to investigate the effect of noise on the dynamics of the qubit. With the above approximations, the non-unitary evolution of a qubit, i.e., evolution including the effects of the environment such as relaxation and dephasing can be governed by the Lindblad master equation. This equation is the most general trace-preserving form of the evolution for the reduced density matrix of the qubit $\rho_{\rm q}$ (means that the sum of probabilities over any complete set of orthogonal states is one) and guarantees the positivity of the reduced density matrix of the qubit at all times, i.e., $\langle \psi \vert \rho_{\rm q} (t) \vert \psi \rangle \geq 0 $.

The Lindblad master equation is defined as \cite{Schlosshauer1,Breuer,Manzano,Lindblad}

\begin{align}
\frac{\D \rho_{\rm q}(t)}{\D t}=-i[H_{\rm q}(t),\rho_{\rm q}(t)]+\sum_i \Gamma_i \Big(L_i \rho_{\rm q} L^{\dagger}_i -\frac{1}{2}\big \lbrace L_i L^{\dagger}_i, \rho_{\rm q}\big \rbrace\Big),
\end{align}
where $\lbrace...\rbrace$ shows the anticommutator which defined for $A$ and $B$ operators as $\lbrace A,B \rbrace =AB+BA$. Moreover, $L_i$ are the Lindblad operators or Lindblad generators representing the interaction between the qubit and the environment and $\Gamma_i$ is the time-independent coefficients that encapsulate all information about the physical parameters of the decoherence process. In the Lindblad master equation in Eq. (13), the first term denotes the unitary evolution of the qubit based on the Liouville-von Neumann equation and the second term is a non-unitary term and arises from the effect of the environment. We suppose $\hbar=1$ in the rest of the paper.

To describe the dynamics of superconducting qubit interacting its environment, we need to choose the Lindblad operator $L_i$ based on the studied model. The effect of noise on a single qubit can be modeled by the Lindblad operators as \cite{Kwon}

\begin{subequations}
\begin{align}
L_1=\sqrt{\Gamma_1} \sigma_-  \\
L_2=\sqrt {\Gamma_{\phi}/2}  \sigma_z,
\end{align}
\end{subequations}
where $L_1$ represents the relaxation process, i.e., the transition from $\vert 1 \rangle $ to $\vert 0\rangle$, and $L_2$ shows the dephasing process. 

Considering the relaxation and dephasing processes, the Lindblad master equation for the qubit takes the form \cite{Rasmussen}

\begin{align}
\frac{\D \rho_{\rm q}(t)}{\D t}=-i[H_{\rm q}(t),\rho_{\rm q}(t)]+\Gamma_1\Big(\sigma_- \rho_{\rm q} \sigma_+ -\frac{1}{2} \lbrace \sigma_ -\sigma_+,\rho_{\rm q}\rbrace \Big)+\Gamma_{\phi}  \Big( \sigma_z \rho_{\rm q} \sigma_z-\rho_{\rm q}\Big).
\end{align}
where the decoherence rates, $\Gamma_i$ can be found in Eq. (5).

Note that Eq. (15) presents a general phenomenological form of the Lindblad master equation for a single qubit experiencing energy relaxation and pure dephasing. More specific forms of the Lindblad equation are often derived for particular physical setups, like a qubit coupled to a cavity resonator which then interacts with a thermal bath. In such cases, the effective qubit relaxation rate can be expressed in terms of system parameters like the qubit-cavity coupling strength, cavity decay rate, and detuning \cite{Blais3}. 

The Lindblad master equation in Eq. (15) can be solved numerically by Quantum Toolbox in Python (QuTiP) which is an open source computational physics software library for simulating open quantum systems \cite{Johansson,Nation}.

\subsection{Bloch-Redfield Master Equation}

\subsubsection{Framework}

Any master equation to describe the behavior of an open quantum system that exchanges energy and information (not particles) with the environment should satisfy three basic requirements \cite{Whitney}:

\subparagraph {i.} The Hermiticity of $\rho$ that causes all probabilities to be real.

\subparagraph{ii.} The trace-preserving of $\rho$ that causes the sum of probabilities over any complete set of orthogonal states to one and all probabilities lie between zero and one. 

\subparagraph {iii.} The positivity of $\rho$ at all times means that the probability of all possible states is positive. 

In general, there are two approaches to derive master equations \cite{Breuer}:

\subparagraph{1- Phenomenological approach} It is assumed that the evolution of the system is translationally invariant in time, such as Markovian evolution in Lindblad master equations \cite{Lindblad}. It is proved that the Lindblad equation is the most general equation that satisfies the first two conditions above. On the other hand, it also preserves complete positivity which is stronger than positivity, so the third condition is also satisfied.

\subparagraph{2- Perturbative approach} The evolution of a system and its environment is taken from the total Hamiltonian and traces over the degrees of freedom of the environment. Although various methods have been introduced in the approach of perturbative analysis of the interaction between the system and the environment such as Bloch-Redfield \cite{Bloch,Redfield}, Nakajima-Zwanzig \cite{Nakajima,Zwanzig}, Schoeller-Schon \cite{Schoeller}. Under certain approximations, such as the weak coupling limit and the Markov approximation, the Nakajima-Zwanzig formalism can be shown to lead to a master equation of the Redfield form. The relationships between these various perturbative approaches are nuanced, often depending on the specific approximations made in their derivation.

The Bloch-Redfield master equation is a Markovian master equation that describes the time evolution of $\rho$ for a quantum system that is weakly coupled to its environment \cite{Bloch,Redfield}. It is a perturbative master equation to analyze of weak interaction of the systems with varying energies and eigenstates with their environment. The advantage of this master equation is that the decoherence rates can be obtained directly from the features of the environment. However, there is no guarantee that the master equation will always preserve the physical properties of $\rho$, because it is a perturbative approach. Therefore, the Bloch-Redfield master equation should be used carefully. The Lindblad master equation is more reliable because it always produces a valid $\rho$, even though some of its operators may not have a physical justification \cite{Johansson,Nation}.

The Bloch-Redfield master equation is closely related to the original Lindblad master equation. Any Bloch-Redfield master equation transforms a Lindblad equation if a so-called secular approximation (or RWA) is made, where only certain resonant interactions with the environment are preserved. However, unlike Lindblad master equations, they do not guarantee the positivity of $\rho$ at all times. It means that it is possible to get negative populations during the time evolution of $\rho$.

The Bloch-Redfield master equation has been, and continues to be, used extensively to study the dissipative behavior of qubits interacting with their environment.\cite{Whitney}. In the following, we analyze the formalism to derive the Bloch-Redfield master equation for a superconducting qubit coupled to the environment which is a very important approach in CQED.

\subsubsection{Derivation of the Bloch-Redfield master equation}

The starting point of the Bloch-Redfield master equation for the qubits and their environment is the total Hamiltonian in Eq. (10). The most general form of a master equation for the qubit is calculated by tracing out the environmental degrees of freedom from the Liouville-von Neumann relation in Eq. (9) for the total system. In the interaction picture, we obtain

\begin{align}
\frac{\D}{\D t}\rho_{\rm q}(t) = - \int_0^t \D \tau {\tr}_\varepsilon \big[H_{\Int}(t), [H_{\Int}(\tau), \rho_{\rm q}(\tau)\otimes\rho_{\rm \varepsilon}]\big],
\end{align}
where we consider the Born approximation as $\rho(t) \approx \rho_{\rm q}(t) \otimes \rho_{\rm \varepsilon}$. It indicates that there is no entanglement between the qubit and its environment at any time during its evolution. As we have mentioned, the Born approximation justified for weak interaction of the qubit and the environment.

The master equation in Eq. (16) is non-Markovian which means the change in $\rho_{\rm q}$ at time $t$ depends on states at all times $\tau< t$ making it intractable to solve. Therefore, we introduce the master equation in the Markovian regime in which $\rho_{\rm q}(\tau)$ is replaced by $\rho_{\rm q}(t)$ and we extend the integration to infinity and substitute $\tau\rightarrow t-\tau$. The Markov approximation is thus justified if the time scale over which the qubit state changes appreciably is large compared to the time scale over which the environmental correlation functions decay $\tau_ {\rm decay}$. Therefore, we have successfully achieved the Markovian quantum master equation as \cite{Breuer}

\begin{align}
&\frac{\D}{\D t}\rho_{\rm q}(t) = - \int_0^\infty \D \tau\; {\rm Tr}_{\rm \varepsilon} \big[H_{\Int}(t), [H_{\Int}(t-\tau), \rho_{\rm q}(t)\otimes\rho_{\rm \varepsilon}]\big],
\end{align}
This equation is the Redfield equation which is local in time with respect $\rho_{\rm q}(t)$ \cite{Redfield}.

To study a specific case, we assume that the interaction of the qubit and the environment is the form $H_{\Int} = \sum_\alpha A_\alpha \otimes B_\alpha$ where $A_\alpha$ and $B_\alpha$ are the qubit and the environmental operators, respectively and $\alpha$ denotes interaction terms. So, we have a master equation based on the qubit operators and the environmental correlation functions as

\begin{align}
\frac{\D}{\D t} \rho_{\rm q}(t)&=-\sum_{\alpha \beta}\int_0^{\infty} \D \tau \Big\lbrace g_{\alpha \beta} (\tau) \Big[ A_{\alpha} (t) A_{\beta} (t-\tau)\rho_{\rm q}(t) -A_{\alpha} (t-\tau)\rho_{\rm q}(t)A_{\beta} (t)\Big]\nonumber\\
& \times g_{\alpha \beta} (-\tau)\Big[\rho_{\rm q} (t)A_{\alpha} (t-\tau) A_{\beta} (t)-A_{\alpha} (t)\rho_{\rm q}(t)A_{\beta} (t-\tau)\Big]\Big\rbrace ,
\end{align}
where the environmental correlation functions are

\begin{align}
g_{\alpha\beta}(\tau) = {\rm Tr}_{\rm \varepsilon}\left[B_\alpha(t)B_\beta(t-\tau)\rho_{\rm \varepsilon}\right] = \left<B_\alpha(\tau)B_\beta(0)\right>, 
\end{align}
since $\rho_{\rm \varepsilon}$ is a steady state.

In the eigenstates of $H_{\rm q}$, where $A_{mn}(t) = A_{mn} e^{i\omega_{mn}t}$, $\omega_{mn} = \omega_m - \omega_n$ and $\omega_m$ is the frequency of the state $\vert m \rangle$, we obtain

\begin{align}
\frac{\D}{\D t}\rho_{ab}(t) &=-i\omega_{ab}\rho_{ab}(t)-\sum_{\alpha,\beta} \sum_{c,d}^{\rm sec} \int_0^\infty \D \tau \Big\lbrace g_{\alpha \beta} (\tau) \Big[\delta_{bd}\sum_n A^\alpha_{an}A^\beta_{nc} e^{i\omega_{cn}\tau} - A^\alpha_{ac}A^\beta_{db}e^{i\omega_{ca}\tau}\Big]\nonumber\\
&+g_{\alpha \beta} (-\tau) \Big[\delta_{ac}\sum_n A^\alpha_{dn}A^\beta_{nb} e^{i\omega_{nd}\tau} - A^\alpha_{ac}A^\beta_{db}e^{i\omega_{bd}\tau}\Big]\Big\rbrace\rho_{cd}(t),
\end{align}
where "sec" denotes the secular terms which satisfy $|\omega_{ab}-\omega_{cd}| \ll \tau_ {\rm decay}$. 

In the next step, we rewrite $g(\tau)$ in terms of the noise-power spectrum of the environment $S(\omega) = \int_{-\infty}^\infty \D\tau e^{i\omega\tau} g(\tau)$ as

\begin{align}
\int_0^\infty \D\tau\; g_{\alpha\beta}(\tau) e^{i\omega\tau} = \frac{1}{2}S_{\alpha\beta}(\omega) + i\lambda_{\alpha\beta}(\omega),
\end{align}
where $\lambda_{ab}(\omega)$ is an energy shift that we ignore.

After some calculations, the final form of the Bloch-Redfield master equation is \cite{Johansson,Nation}

\begin{align}
\frac{\D}{\D t}\rho_{ab}(t)=-i\omega_{ab}\rho_{ab}(t)+\sum_{c,d}^{\rm sec} R_{abcd}\rho_{cd}(t),
\end{align}
where $R_{abcd}$ is the Bloch-Redfield tensor. For simplicity, we assume that $A_{\alpha}$ are Hermitian and there are no correlations between different environmental operators. Therefore, the final form of the Bloch-Redfield tensor is \cite{Johansson,Nation}

\begin{align}
R_{abcd} =  -\frac{1}{2} \sum_{\alpha \beta}\Big\lbrace\delta_{bd}\sum_nA^\alpha_{an}A^\beta_{nc}S_{\alpha \beta}(\omega_{cn})- A^\alpha_{ac} A^\beta_{db} S_{\alpha \beta}(\omega_{ca})+\delta_{ac}\sum_n A^\alpha_{dn}A^\beta_{nb} S_{\alpha \beta}(\omega_{dn})- A^\alpha_{ac}A^\beta_{db} S_{\alpha \beta}(\omega_{db})\Big\rbrace.
\end{align}

For instance, when modeling a transmon qubit where 1/f flux noise is a dominant dephasing mechanism at its optimal point, one might use the Bloch-Redfield formalism with an appropriate flux noise spectral density as input to Eq. (23).

The Bloch-Redfield master equation in Eq. (22) can be solved numerically using QuTiP. The input parameters are $H_{\rm q}$,  $A_{\alpha}$, and  $S_{\alpha \beta}$ which can be defined based on the type of the qubit-environment interaction.

While the Bloch-Redfield equation offers a more direct connection to the microscopic environment properties than the phenomenological Lindblad approach, care must be taken as it does not always guarantee positivity, unlike the Lindblad form which can be achieved via a secular approximation. Moreover, while Markovian master equations like Lindblad and Bloch-Redfield are widely applicable, situations involving strong system-environment coupling or structured environments with memory necessitate non-Markovian approaches, which we discuss next.

\subsection{Non-Markovian Master Equation}

\subsubsection{Non-Markovian regime}

The Lindblad master equation is a common approach to analyzing the decoherence effects of the superconducting qubit as an open quantum system. This equation is completely positive and derived based on the Born-Markov approximations. These assumptions account the weak qubit-environment interaction and memoryless feature of the environment [3, 4]. Markovian approximation leads to significant simplifications, however in a real physical situation such as in a low-temperature environment and strong interactions of the system with its environment, we should consider the non-Markovian effects \cite{Schlosshauer1,Breuer}. A specific example of non-Markovian dynamics is a superconducting qubit that strongly interacts with a low-temperature environment. In this case, it is not appropriate to use the Born-Markov approximations anymore \cite{Prokofev,Dube}.

Non-Markovian effects may arise from imperfect control and calibration of the qubit \cite{Jeroen}. These effects can also be caused by spatially correlated noise originating from non-local pulse controls, coherent errors caused by residual Hamiltonian terms and stochastic errors due to slow environmental fluctuations. The non-Markovian effects can be coherent, due to unwanted interaction with neighboring qubits, or incoherent, due to coupling with magnetic impurities \cite {Paladino, Zhang}.

The witnesses of non-Markovianity behavior of a quantum system can be classified based on \cite{Rivas}:

\subparagraph{i.} monotonicity under completely positive maps such as quantum relative entropies, quantum Fisher information, fidelity, Bloch volume measure and capacity measures. 

\subparagraph{ii.} monotonicity under local completely positive maps including entanglement, quantum mutual information, and quantum discord. 

Moreover, some measures of non-Markovianity are introduced to quantify the non-Markovianity effects of a system, especially in real experiments. A measure of non-Markovianity is a function that determines the Markovian and non-Markovian dynamics of a process based on zero and positive numbers, respectively. In this regard, some measures of quantum non-Markovianity are proposed such as geometric measures, Helstrom matrix norm, decay rates measures and RHP (Rivas, Huelga and Plenio) measure. For more details, please see \cite{Rivas}.

The non-Markovian noise is an important challenge in quantum error correction \cite{Landahl,Fowler,Devitt}, fault-tolerant quantum computation \cite{Alicki,Aharonov,Ng} and other near-term quantum technologies \cite{Sarovar}. In particular, non-Markovian effects can be considered an important source of errors in superconducting qubits, leading to errors that come in the form of unwanted terms in the qubits' Hamiltonian and lead to unwanted operations \cite{Kohout}. In superconducting qubits, the two main non-Markovian signatures are the periodic revival of coherence and the appearance of additional frequencies far from the qubit frequency in the qubit state precession \cite{Gulaci}. However, the methods commonly employed to test and confirm devices in quantum computing are not able to fully measure non-Markovian effects. This is because these methods assume that the Markovian approximation is true, either directly or indirectly. However, recent studies have observed deviations of qubit dynamics based on the Markov approximation \cite{Rudinger,Tripathi,Pokharel,White}. As a result, in situations where noisy processes cannot be analyzed using normal Markov methods, it seemed to be very important to analyze the dynamics of the qubits such as superconducting qubits in the non-Markovian regime. 

In the next subsection, we describe a master equation to model the noise in the superconducting qubit based on the non-Markovian dynamics.

\subsubsection{Non-Markovian dynamics of superconducting qubits}

In recent years, some studies have been done to analyze the non-Markovian dynamics of qubits, especially superconducting qubits \cite{Zhang,Gulaci,White,Niu,Tancara,Chen1,Banchi}. For example, one of the first and most accurate non-Markovian master equations is the Nakajima–Zwanzig (NZ) master equation which can be defined as \cite {Breuer,Banchi}

\begin{align}
\frac{\partial \rho (t)}{\partial t}=-i [H,\rho(t)]+\int_0^t \D s \mathcal{K}^{\NZ}_{t-s}[\rho(s)] + \chi(t),
\end{align}
where $\mathcal{K}^{\NZ}_{t-s}$ is a superoperator called memory kernel that describes the interaction of the qubit with the environment. Moreover, $\chi(t)$ shows the initial correlations between the qubit and the environment. So, if the qubit and environment are initially uncorrelated, we will have $\chi(t)=0$. The NZ master equation describes non-Markovian dynamics because the state at time $t+\D t$ depends not only on $\rho(t)$  but also on the states $\rho(s)$ for $s<t$. However, in the time domain, the NZ equation is not easy to solve numerically. Therefore, an equally accurate master equation has been proposed called the time-convolutionless (TCL) master equation \cite {Breuer,Banchi}

\begin{align}
\frac{\partial \rho (t)}{\partial t}=-i [H,\rho(t)]+\int_0^t \D s \mathcal{K}^{\TCL}_{t-s}[\rho(t)] + \chi(t),
\end{align}

The NZ master equation is different from the TCL. In the NZ case, the whole history of states is considered, while in the TCL case, only the current state $\rho(t)$ is taken into account and non-Markovian effects are included in the memory kernel. The main problem in the TCL mater equation is that the interaction with the environment is normally unknown and the memory kernel can be obtained using perturbation theory \cite {Banchi}.

To analyze the non-Markovian noise effects on superconducting qubits, we focus on the post-Markovian master equation (PMME) which has the form \cite{Shabani}

\begin{align}
\frac{\D}{\D t}\rho(t)= L_0 \rho(t)+L_1 \int_0^t \D t' \mathcal{K}(t') \EXP\big[{({L_0+L_1})t'} \big]\rho(t-t'),
\end{align}
where $\rho(t)$ is the reduced state of the qubit and $L_0$ and $L_1$ are the Lindbladian generators corresponding to the Markovian and non-Markovian dynamics, respectively.  Moreover, the memory kernel $\mathcal{K} (t)$ assigns weights to the previous history of the system states. The master equation in Eq. (26) is derived phenomenologically, because it is used to interpolate between Markovian and exact dynamics in the measurement interpretation \cite{Shabani}. In the limit $\mathcal{K}(t) \mapsto\delta(t)$, the Markovian dynamics is obtained \cite{Zhang,Agarwal}.

In the PMME, we consider both noise in the Markovian and non-Markovian regimes. The Lindbladian generators for a superconducting qubit in the PMME can be defined as \cite{Agarwal}

{\bf (i)} $L_0$ shows the Markovian Lindblad operators that include amplitude damping and dephasing. 

{\bf (ii)} $L_1$ shows the Lindbladian operators due to non-Markovian dephasing that is written as 

\begin{align}
L_1\rho(t)=\gamma_z\big((\sigma_z\rho(t) \sigma_z-\rho(t)\big).
\end{align}
where $\gamma_z$ quantifies the amount of non-Markovian dephasing \cite{Agarwal}. To solve the PMME, we assume the memory kernel has the form $\mathcal{K}= \EXP({-\gamma t})$, in which $\gamma$ is the decay rate of the memory kernel  \cite{Zhang,Shabani}. However, more general forms of $\mathcal{K}(t)$ can be used to analyze the wider range of non-Markovian dynamics of superconducting qubits. 

In general, we can use the PMME  to model the non-Markovian effects in superconducting qubits for its computational simplicity, analytical solvability and retain complete positivity with an appropriate form of $\mathcal{K} (t)$ \cite{Chruscinski,Sutherland}.

In the next section, we use an approach based on an effective Hamiltonian to describe the non-Hermitian dynamics of open quantum systems.

\subsection{Monte Carlo Approach}

\subsubsection{Framework}

It is usually assumed in quantum mechanics that the Hamiltonian describing a quantum system is Hermitian, thus guaranteeing the reality of energy eigenvalues. However, the dissipative behavior of open quantum systems is described by non-trace-preserving and non-Hermitian Hamiltonians \cite{Naghiloo}. In this context, effective Hamiltonian methods were developed to better understand the quantum measurement theory \cite{Chen2}. Moreover, some experiments with photons \cite{Xiao,Klauck,Yu}, spins \cite{Wu,Liu} and superconducting qubits \cite{Naghiloo} represent special quantum effects in non-Hermitian dynamics.

So far, various approaches have been presented for the analysis of the dynamics of open quantum systems and decoherence theory \cite{Schlosshauer1,Breuer}. In this regard, the Monte Carlo wave function method (quantum jump method or quantum trajectory method) is a computational technique to simulate open quantum systems and quantum dissipation was developed in the early 1990s \cite{Dum,Molmer,Plenio,Gardiner}. The Monte Carlo is an approach that is very similar to the methods based on the master equations, with the difference that it operates on the wave function instead of using the density matrix. The basis of this method is the time evolution of the wave function of the system with an effective Hamiltonian, so that at each time step, a discontinuous change (quantum jump) may occur with a small probability. The quantum trajectory can be obtained by the calculated state of the system $\psi(t)$  and the density matrix $\rho(t)$ can be obtained by averaging over many trajectories \cite{Dum,Molmer}.

The Monte Carlo method has two parts: First, the time evolution of the system with a non-Hermitian Hamiltonian. Then, random quantum jumps are made, and the wave function is adjusted to account for these changes. Moreover, there are two main features of this method. First, if the dimension of the Hilbert space of the quantum system is $N$, the number of variables involved in a method based on wave function ($\approx N$) is much smaller than the amount required for the method based on density matrices ($\approx N^2$). Consequently, Monte Carlo is a useful theoretical method to reduce the dimensionality of the numerical problem. Second, we can gain a new physical insight into how a single quantum system behaves \cite{Dum,Molmer,Kornyik}

Recently, the Monte Carlo approach has been used to investigate the effects of noise on superconducting qubits \cite{Gelman,Ruggiero,Kerman,King,Kiss,Mickelsen,Ozfidan,Sett}. In the next subsection, we formulate the procedure for evolving wave functions of the superconducting qubits affected by environmental noises based on the Monte Carlo technique.

\subsubsection{Formalism of the Monte Carlo approach}

As we discussed in section A, a master equation for a qubit interacting with noise based on the Born-Markov approximation in the most general form -Lindblad- can be written as \cite{Schlosshauer1,Breuer}

\begin{align}
\frac{\D \rho_{\rm q}(t)}{\D t}&=-i[H_{\rm q}(t),\rho_{\rm q}(t)]+\sum_m \Gamma_m \Big(C_m \rho_{\rm q} C^{\dagger}_m -\frac{1}{2}\big \lbrace C_m C^{\dagger}_m, \rho_{\rm q}\big \rbrace\Big) \nonumber\\
& \equiv 2 Re \Big \lbrace  \frac{H_{\rm eff} \rho}{i} \Big \rbrace +\sum_m C_m \rho C^{\dagger}_m,
\end{align}
where $H_{\rm eff}$ denotes a non-Hermitian effective Hamiltonian which defined as

\begin{align}
H_{\rm eff}=H_{\rm q}-\frac{i}{2}\sum_{m}C^{+}_{m}C_{m},
\end{align}
in which $C_m$ are quantum jump or collapse or Lindblad operators and their maximum number is one less than the square of the dimension of the system. The evolution of the qubit's state $\vert\psi(t)\rangle$ interacting with noises is governed by the Schrödinger equation with $H_{\rm eff}$ \cite{Molmer,Kornyik}. In its original form, the Monte Carlo method to evolve $\vert\psi(t)\rangle$ to $\vert\psi(t+\delta t)\rangle$ can be defined as follows.

The effective Hamiltonian gives the new wave function of qubit at sufficiently small $\delta(t)$ obtained by evolving $\vert\psi(t)\rangle$

\begin{align}
\vert\psi^{(1)}(t+\delta t) \rangle= \Big(1-iH_{\rm eff}\delta(t)\Big) \vert \psi(t)\rangle,
\end{align}
Since $H_{\rm eff}$ is non-Hermitian, the new wave function is not normalized. Therefore, the norm of the wave function, that to first-order in $\delta(t)$ can be written as 

\begin{align}
&\left<\psi^{(1)}(t+\delta t) \vert \psi^{(1)}(t+\delta t) \right> =\langle \psi(t) \vert \Big(1+iH_{\rm eff}^{\dagger}\delta(t)\Big)\Big(1-iH_{\rm eff}\delta(t)\Big) \vert \psi(t)\rangle =1-\delta p,
\end{align}

where 

\begin{subequations}
\begin{align}
&\delta p =i\delta t  \langle \psi(t) \vert H_{\rm eff}-H_{\rm eff}^{\dagger} \vert \psi(t) \rangle=\sum_m \delta p_m \\
&\delta p_m =\delta t \left<\psi(t)|C^{+}_{m}C_{m}|\psi(t)\right> \geq 0,
\end{align}
\end{subequations}

Note that the magnitude of the time-step $\delta(t)$ should be small enough for the first-order calculation to be valid; in particular, it requires $\delta p \ll 1$ \cite{Molmer,Kornyik}.

The next step of the evolution of $\psi(t)$ between $t$ and $t + \delta t$ consists of a possible quantum jump. If environmental measurements detect a quantum jump, for example through the emission of a photon into the environment, the wave function of qubit undergoes a jump into a state defined by projecting $\vert \psi(t)\rangle$ using $C_{m}$ operators corresponding to the measurement theory \cite{Molmer}. In other words, the environment is continuously monitored and this makes the system's wave function jump in certain ways. These quantum jumps depend on how much information is gained about the system through the measurements of the environment.

To determine whether a quantum jump occurs, we use a random number $0 < \epsilon < 1$. If $\delta_p < \epsilon$ (a usual case since $\delta_p\ll 1$), no quantum jump happens. In this case, we use the following relation to find the new normalized wave function of the qubit

\begin{align}
\vert \psi(t+\delta t)\rangle=\vert\psi^{(1)}(t+\delta t)\rangle / (1-\delta p)^{1/2},
\end{align}
If $\epsilon < \delta p$, then a quantum jump happens. We obtain the new normalized wave function from the different states $C_m \vert \psi (t)\rangle$ with the probability distribution

\begin{align}
 \Pi_m=\delta p_m / \delta p, 
\end{align}
where $\sum_m \Pi_m=1$.  As a result, the normalized wave function is \cite{Molmer}

\begin{align}
\left|\psi(t+\delta t)\right>=C_{m}\left|\psi(t)\right>/\big(\delta p_m/\delta t)^{1/2},
\end{align}
Note this condition is important that the probability of two quantum jumps occurring at the same time is negligible.

It can be easily shown that the above steps in the time evolution of the qubit wave function $\vert \psi(t)\rangle$ are equivalent to the master equation to first order in $\delta t$ \cite{Molmer,Kornyik}. Considering Eqs. (33) and (35), the density matrix of the qubit can be written as

\begin{align}
\rho(t+\delta t) &= \vert \psi(t+\delta t) \rangle \langle \psi(t+\delta t)\vert \nonumber\\
&=(1-\delta p) \vert \psi(t+\delta t) \rangle \langle \psi(t+\delta t)\vert \Big\vert_{\rm no-jump}+\sum_m \delta p_m \vert \psi(t+\delta t) \rangle \langle \psi(t+\delta t)\vert \Big\vert_{\rm mth-jump}\nonumber\\
&=\rho(t)+\delta t \Big(\frac{H_{\rm eff}}{i } \rho (t)-\rho(t) \frac{H^{\dagger}_{\rm eff}}{i } \Big)+\delta t \sum_m C_m \rho(t) C^{\dagger}_m +\mathcal{O} (\delta t^2).
\end{align}

The Monte Carlo method tries to unravel a master equation into different possible quantum trajectories of the system using random numbers. Each trajectory starts with a specific set of conditions that represent the different possible states of the system \cite{Kornyik}.

To simulate the dynamics of the superconducting qubits interacting with noises using the Monte Carlo method, we use QuTiP based on the following algorithm \cite{Nation}:

\subparagraph{i.} Starting from a pure state $\vert \psi(0)\rangle$, we select a random number $\epsilon$ between zero and one. This number represents the probability of a quantum jump happening.

\subparagraph{ii.} We solve the Schrödinger equation using the $H_{\rm eff}$ in Eq. (29) until $\langle \psi (\tau) \vert  \psi (\tau) \rangle =\epsilon$ that $\tau$ is a time in which a quantum jump occurs.

\subparagraph{iii.} The resultant jump projects the qubit's state at $\tau$ into one of the renormalized states given by Eq. (35). The corresponding jump operator $C_m$ is chosen so that $m$ is the smallest possible integer to satisfy the following relation

\begin{align}
\sum_{i=1}^{m} \Pi_{m}(\tau) \ge \epsilon,
\end{align}

\subparagraph{iv.} We use the renormalized state from step {\bf (iii)} as an initial condition at $\tau$ and select a new $\epsilon$ and repeat the above steps until the simulation is finished.

For a concrete example, in \cite{Chen2}, the Monte Carlo method is used to analyze the non-Hermitian dynamics of a superconducting qubit which is perturbed by quantum jumps between energy levels. In this experiment, the lowest three energy levels $\vert g \rangle$, $\vert e \rangle$ and $\vert f \rangle$ of a transmon superconducting circuit are considered and the transmon circuit is placed within a three-dimensional microwave cavity. The effective Hamiltonian is defined as 

\begin{align}
H_{\rm eff}=H_{\rm c}-iC^{\dagger}_e C_e/2-iC^{\dagger}_f C_f/2.
\end{align}
The jump operatorts are $C_e=\sqrt{\gamma_e} \vert g \rangle \langle e \vert$ and $C_e=\sqrt{\gamma_f} \vert e \rangle \langle f \vert$ in which $\gamma_e$ and $\gamma_f$  are the decay rate of $\vert e \rangle $ and $\vert f \rangle$, respectively. Moreover, $H_c$ is

\begin{align}
H_c=J  \big( \vert e \rangle \langle f \vert +\vert f \rangle \langle e \vert \big) +\frac{\Delta}{2} \big( \vert e \rangle \langle e \vert +\vert f \rangle \langle f \vert \big),
\end{align}
where $\Delta$ is the frequency detuning of $\vert e \rangle- \vert f \rangle $ transition and $J$ is the coupling strength of the transmon with the microwave cavity. This study shows the role of quantum jumps in generalizing the applications of classical non-Hermitian systems to superconducting qubits as open quantum systems for sensing and control.

\subsection{Floquet Theory}

The effect of an oscillating field on the superconducting qubits causes changes in qubits's macroscopic quantum state \cite{Vion,Chiorescu}. Therefore, the analysis of the dynamics of qubits interacting with the strong oscillating field is important to quantum computing based on superconductivity. For example, superconducting qubits usually have a short coherence time \cite{Martinis2,Martinis3}. Therefore, a strong AC field is often used to diminish each gate operation time \cite{Son}. Consequently, we need to develop an approach for a time-dependent problem in superconducting qubits.

A lot of time-dependent problems in physics are periodic. We can describe the dynamics of these systems by numerical methods. They can also be transformed into time-independent problems that can be solved much more efficiently. In this regard, the Floquet theory is a very useful tool to do this transformation \cite{Grifoni,Chu}. This theory has been developed for some time-dependent problems such as atomic and molecular systems \cite{Chu,Chu2,Telnov} and laser fields \cite{Son2}. The Floquet method is based on a non-perturbative analysis of the interaction of a quantum system with intense time-dependent fields \cite{Son}.

Recently, Floquet theory has attracted much attention in superconducting qubits-based technologies. For example, the Floquet dynamics of two weakly coupled flux qubits is investigated under strong periodic driving with different conditions \cite{Song}. We can also use the properties of periodic Hamiltonians to increase the dephasing time of qubits \cite{Gandon}. Moreover, it is shown that the control of qubits using periodic force, called Floquet engineering, leads to the qubit being governed by a time-independent Floquet Hamiltonian with new properties \cite{Petiziol}. In an interesting study, a new approach is proposed to reduce the susceptibility of a qubit to $1/f$ noise using a periodic drive field. This study introduces a method to encode quantum information in robust and time-dependent states which can be considered the new approach for quantum information processing \cite{Huang}. Therefore, it seems that the Floquet theory is an important model for analyzing the effects of noise, control, and manipulation of superconducting qubits \cite{Chen3}.

In the following subsections, we first formulated the general formalism of Floquet theory. Then, we provide a Floquet master equation to analyze the dynamics of superconducting qubits in the presence of environmental noises.

\subsubsection{Floquet formalism}

Let us start with the time-dependent Schrödinger equation 

\begin{align}
H(t) \vert \Psi(t)\rangle = i\frac{\partial}{\partial t} \vert \Psi(t)\rangle,
\end{align}
where $H(t)$ and $\vert \Psi(t) \rangle$ are the Hamiltonian and wave function of the system, respectively.

In the presence of intense fields interacting with the system, we are interested in the analysis of the dynamics of quantum systems that their Hamiltonian are a periodic function in time

\begin{align}
H(t) = H(t+T),
\end{align}
where $T$ is the period of the perturbation. Considering the symmetry of the Hamiltonian under time translation, $t\rightarrow t+T$, we can use the Floquet formalism to describe the dynamics of the system \cite{Grifoni,Floquet,Ince,Magnus}.

In the Floquet theory, the solutions of Eq. (40)-called Floquet states- have the general form

\begin{align}
\vert \Psi_\alpha(t)\rangle = \exp(-i\varepsilon_\alpha t) \vert \Phi_\alpha(t)\rangle,
\end{align}
where $\vert\Phi_\alpha(t)\rangle$ is a Floquet mode that is periodic in time

\begin{align}
\vert\Phi_\alpha(t)\rangle=\vert\Phi_\alpha(t+T)\rangle,
\end{align}
Moreover, $\varepsilon_\alpha$ is a real-valued energy coefficient that is termed the Floquet characteristic exponent or the quasienergy levels. They are constants in time, but only uniquely defined up to multiples of $ \Omega$ in the interval $[0,2\pi/T]$ where $\Omega=2\pi/T$.

Knowing the Floquet modes $\vert\Phi_\alpha(t)\rangle$ at $t \in [0,T]$ and the quasienergies $\epsilon_\alpha$ for a specific $H(T)$, we can write any initial wave function $\vert\Psi (0)\rangle$ into Floquet states and obtain the solution at $t$ as

\begin{align}
\vert\Psi(t)\rangle = \sum_\alpha c_\alpha \vert\Psi_\alpha(t) \rangle= \sum_\alpha c_\alpha \exp(-i\varepsilon_\alpha t)\vert\Phi_\alpha(t)\rangle,
\end{align}
where the coefficients $c_\alpha$ can be calculated by the initial wave function $\vert\Psi(0) \rangle= \sum_\alpha c_\alpha \vert\Psi_\alpha(0)\rangle$. By substituting Floquet states, Eq. (42), into the Schrödinger equation, Eq. (40), we obtain an eigenvalue problem for the Floquet modes and quasienergies as

\begin{align}
\mathcal{H}(t)\vert\Phi_\alpha(t)\rangle = \varepsilon_\alpha\vert\Phi_\alpha(t)\rangle,
\end{align}
where Floquet Hamiltonian $\mathcal{H}(t)$ is

\begin{align}
\mathcal{H}(t) = H(t) - i \frac{\partial}{\partial_t},
\end{align}
Note that $\vert\Phi_{\alpha,n}(t)\rangle\equiv\exp(-in\Omega t) \vert\Phi_\alpha(t)\rangle$ yields a solution of Eq. (40) which is physically identical to Eq. (42) but with the shifted quasienergy $\varepsilon_{\alpha,n}\equiv \varepsilon_\alpha-n \Omega$. Moreover, $n=0,\pm 1, \pm 2, ...$ is an integer number, $\alpha$ denotes a whole class of solutions,  $\varepsilon_\alpha=\varepsilon_{\alpha,0}$ and $\vert\Phi_\alpha(t)\rangle=\vert\Phi_{\alpha,0}(t)\rangle$. Therefore, the eigenvalues $\lbrace \varepsilon_{\alpha,n} \rbrace $ can be mapped into a first Brillouin zone in which $- \Omega/2 \leq \varepsilon_{\alpha,n} \leq \Omega/2$ \cite{Grifoni,Hausinger}.

For $T$-periodic functions ($f,g$), we introduce the Hilbert space ${\mathcal I}$ in which the inner product of the functions is 

\begin{align}
(f,g)=\frac{1}{T} \int_0^T \D t f^* (t) g(t),
\end{align}

In the Hilbert space ${\mathcal I}$, the functions $\phi_n(t)=\exp(-in\Omega t)$ construct a complete and orthonormal basis set \cite{Simmons}. Moreover, we define a basis-independent notation for the state vectors $\vert n )$ where $\phi_n(t)=(t \vert n )$ \cite{Hausinger}. In the composite Hilbert space of the Floquet Hamiltonian, ${\mathscr H} \otimes {\mathcal I}$, the scalar product can be defined as 

\begin{align}
\langle \langle . \vert . \rangle \rangle=\frac{1}{T} \int_0^T \D t \langle . \vert . \rangle,
\end{align}
We can write a $T$-periodic state vector $\vert\Phi_{\alpha,n}\rangle$ in ${\mathscr H}$ as a Fourier series and expand it in basis functions of ${\mathcal I}$

\begin{align}
\vert\Phi_{\alpha,n}(t)\rangle=\exp(-in\Omega t) \vert\Phi_\alpha (t)\rangle=\sum_l \exp(-il\Omega t) \vert\Phi^{(n-l)}_\alpha\rangle,
\end{align}
where $\vert\Phi^{(k)}_\alpha\rangle$ is the time-independent Fourier coefficient. In the Hilber space ${\mathcal H} \otimes {\mathcal I}$, we define

\begin{align}
\vert\Phi_{\alpha,n}\rangle \rangle \equiv \sum_l \vert\Phi^{(n-l)}_\alpha \rangle \otimes  \vert l),
\end{align}

Considering the expansion of the Hilbert space, we can now treat a time-dependent problem in Eq. (40) like a time-independent one \cite{Grifoni,Hausinger}.

To find the Floquet states and quasienergies, we consider the propagator for the Schrödinger equation, Eq.(40), which satisfies

\begin{align}
U(T+t,t)\vert\Psi(t)\rangle = \vert\Psi(T+t)\rangle,
\end{align}
By inserting the Floquet states in Eq. (42) into Eq. (51) and regarding Eq. (43), we have

\begin{align}
U(T+t,t)\vert\Phi_\alpha(t)\rangle = \exp(-i\varepsilon_\alpha T)\vert\Phi_\alpha(t)\rangle = {\mathcal G}_\alpha \vert\Phi_\alpha(t)\rangle,
\end{align}
which represents that the Floquet modes $\vert\Phi_\alpha(t)\rangle$ are the eigenstates of the propagator ${\mathcal G}_\alpha$. Consequently, by numerical calculating and diagonalizing of $ U(T+t,t)$, we can find the Floquet modes and quasienergies. This method is especially useful to find $\vert\Phi_\alpha(0)\rangle$ by diagonalizing $U(T,0)$ \cite{Nation}.

The Floquet modes at arbitrary $t$ can be obtained by propagating $\vert\Phi_\alpha(0)\rangle$ to $\vert\Phi_\alpha(t)\rangle$ using $U(t,0)\vert\Psi_\alpha(0) \rangle= \vert\Psi_\alpha(t)\rangle$, which for the Floquet modes, we have

\begin{align}
U(t,0)\vert\Phi_\alpha(0)\rangle = \exp(-i\varepsilon_\alpha t)\vert\Phi_\alpha(t)\rangle,
\end{align}
Since $\vert\Phi_\alpha(t)\rangle$ is periodic, we only need to calculate it for $t \in [0,T]$ and from $\vert\Phi_\alpha (t \in [0,T])\rangle$, we can directly obtain $\vert\Phi_\alpha(t)\rangle$, $\vert\Psi_\alpha(t)\rangle$ and $\vert\Psi (t)\rangle$ for large $t$ \cite{Nation}.

To analyze the dynamics of the system in the Floquet formalism, we solve the eigenvalue problem in Eq. (45) analytically or numerically by softwares such as QuTiP \cite{Johansson,Nation,Creffield}. QuTiP provides a group of functions to calculate Floquet modes, quasienergies and Floquet mode decomposition considering the time-dependent Hamiltonian. 

The Floquet formalism is only appropriate to find $\vert\Psi(t)\rangle$ of Hamiltonian $H(t)$, if we can evaluate the Floquet states and quasienergies more easily than solving the time-dependent Schrödinger equation in Eq. (40). This formalism is a useful tool to study strongly driven periodic quantum systems. This non-perturbative formalism is a useful tool to study strongly driven periodic quantum systems. It explicitly handles the time-periodic nature of the driving Hamiltonian (which can be thought of as a time-dependent perturbation to the undriven system's dynamics), and by transforming to the Floquet basis, it helps in avoiding secular terms that can arise in naive perturbative treatments of periodically driven systems \cite{Grifoni}.

In the next subsection, we formulated the Floquet master equation to investigate the effects of noises on superconducting qubits.

\subsubsection{Floquet master equation}

To analyze the dynamics of the qubits taking into account the time-dependent Hamiltonian, we need a more rigorous method. In this regard, the Markovian Floquet master equation is provided, with important applications for strongly driven systems such as superconducting qubits \cite{Huang,Chen3,Hausinger}. In this approach, the master equation is obtained based on the Floquet theory \cite{Grifoni,Breuer2,Kohler}. 

Let us describe the detailed framework of the Floquet master equation for superconducting qubits influenced by a periodic drive.

As we discussed in the previous subsection, the quasienergies $\varepsilon_j$ and time-periodic Floquet states $\vert \Phi_j(t)\rangle$ of the driven qubit considered as solutions of the Floquet equation, Eq. (45) in which $\mathcal{H}(t) = H_{\rm q}(t) - i \partial / \partial_t$. In the absence of noise, the driven qubit can be evolved by the time evolution operator as

\begin{align}
 U_{\rm q}(t,0)=\sum_{j=0,1} \vert \Phi_j(t)\rangle \langle \Phi_j(0)\vert \exp (-i \varepsilon _j t),
\end{align}
This evolution leads to the population in each Floquet state remaining invariant and the relative phases accumulate by a rate obtained by the quasienergy difference $\varepsilon_{01}\equiv \varepsilon_1-\varepsilon_0$. 

Considering Floquet theory in open quantum systems, the decoherence time of a driven qubit is defined based on the qubit’s Floquet states \cite{Breuer2,Kohler}. It is shown that the decoherence rate of the driven qubit in the presence of noise can be written as \cite{Huang}

\begin{align}
\gamma_\mu=\int_ {-\infty}^ {\infty} \D \omega F_\mu(\omega)S(\omega),
\end{align}
where $S(\omega)$ is the noise spectrum and $F_\mu(\omega)$ is the filter function in which the effect of noise on a driven qubit in comparison to an undriven one is captured in this function \cite{Martinis4}. In addition, $\mu=\mp,\phi$ represents the different noise channels of relaxation, excitation and pure dephasing of the Floquet qubit, respectively.

The total Hamiltonian of the qubit and the environment can be written as $H(t) = H_q(t) + H_\varepsilon + H_{\Int}$, including the time-periodic qubit Hamiltonian, time-independent environment Hamiltonian and interaction Hamiltonian, respectively.  

Let us consider the Redfield equation of the driven qubit which describes the evolution of the qubit density matrix $\rho_{\rm q}$ in the interaction picture as \cite{Breuer}

\begin{align}
\frac{\D}{\D t}\rho_{\rm q}(t) = - \int_0^t \D \tau\; {\rm Tr}_{\rm \varepsilon} \big[H_{\Int}(t), [H_{\Int}(t-\tau), \rho_{\rm q}(t)\otimes\rho_{\rm \varepsilon}]\big],
\end{align}
Here, $\rho_{\rm \varepsilon}$ is the density matrix of the environment in the interaction picture which is assumed to stay in thermal equilibrium.

Let us suppose the interaction Hamiltonian of the qubit and the environment has the form $H_{\Int}=\sigma \eta$, where $\sigma$ and $\eta$ are the qubit and the environment operators, respectively. The term $H_{\Int}(t)=U^{\dagger}_0(t) H_{\Int} U_0(t)$ which is expressed in the interaction picture. Furthermore, $U_0(t) = U_{\rm q}(t)U_\varepsilon(t)$ where $U_{\rm q}(t)$ is defined in Eq. (54) and $U_\varepsilon(t)=\exp(-i H_\varepsilon t)$. On the other hand, the interaction Hamiltonian can be expressed in the interaction picture as $H_{\Int}(t)=\sigma(t) \eta(t)$, where $\sigma(t)=U^{\dagger}_{\rm q}(t) \sigma U_{\rm q}(t)$ and $\eta(t)=U^{\dagger}_{\varepsilon}(t) \eta U_\varepsilon(t)$. The Redfield equation in Eq. (56) is an integro-differential equation and is not convenient to calculate the decoherence rate $\gamma_\mu$ of the qubit. To obtain $\gamma_\mu$, we should simplify this equation using the RWA. To find the fast-rotating terms, we decompose $\sigma(t)$ into different frequency parts as \cite{Huang}

\begin{align}
\sigma(t)=\sum_{k\in \mathbb{Z},\mu=\pm,\phi}g_{k\mu}c_\mu(0)\exp(-i\omega_{k\mu}t),
\end{align}
where $c_\mu$ denotes the Floquet counterpart of the Pauli matrices

\begin{subequations}
\begin{align}
&c_+(t)=\vert \Phi_1(t) \rangle \langle \Phi_0(t)\vert  \\
&c_-(t)=\vert \Phi_0(t) \rangle \langle \Phi_1(t)\vert \\
&c_\phi(t)=\vert \Phi_1(t) \rangle \langle \Phi_1(t)\vert-\vert \Phi_0(t) \rangle \langle \Phi_0(t)\vert, 
\end{align}
\end{subequations}
Furtheremore, the frequencies $\omega_{k \mu}$ are the filter frequencies defined as $\omega_{k\pm}=\mp \varepsilon_{01}+k\omega_d$ and $\omega_{k\phi}=k\omega_d$ where $k\in \mathbb{Z}$ and $\omega_d$ is the fequency of the driven qubit.

After some manipulations, the Redfield equation can be transformed to \cite{Huang}

\begin{align}
\frac{\D}{\D t}\rho_{\rm q}(t) =\sum_{\mu=\pm,\phi}\zeta_{\mu}\Big[ \int_{-\infty}^{\infty} \D \omega F_\mu (\omega,t)S(\omega)\Big] {\mathcal D}[c_\mu] \rho_{\rm q}(t),
\end{align}
where

\begin{align}
F_\mu(\omega,t)=\frac{1}{\pi \zeta_\mu} \sum_k t \sinc \big[(\omega-\omega_{k\mu})t\big] \vert g_{k\mu}\vert^2,
\end{align}
${\mathcal D}[L] $ is the usual damping superoperator

\begin{align}
{\mathcal D}[L] \rho_{\rm q}(t)=L\rho_{\rm q}L^{\dagger} -(L^{\dagger}L \rho_{\rm q}+\rho_{\rm q}L^{\dagger L})/2,
\end{align}
and noise spectrum is

\begin{align}
S(\omega)=\int_{-\infty}^{\infty}\D t \exp(i\omega t) {\rm Tr}_\varepsilon [\eta(t)\eta(0)\rho_\varepsilon].
\end{align}
Here, $g_{k\mu}$ are the Fourier-transformed coupling matrix elements which are gdefined as

\begin{subequations}
\begin{align}
&g_{k\pm}=\frac{\omega_d}{2\pi} \int_0^{2\pi/ \omega_d} \D t \exp(ik \omega_d t) {\rm Tr}_{\rm q}[\sigma c_{\mp}(t)] \\
&g_{k\phi}=\frac{\omega_d}{4\pi} \int_0^{2\pi/ \omega_d} \D t \exp(ik \omega_d t) {\rm Tr}_{\rm q}[\sigma c_{\phi}(t)].
\end{align}
\end{subequations}
We have further used $\zeta_{\pm}\equiv1$, $\zeta_\phi\equiv 1/2$ and $c_\mu(0)\rightarrow c_\mu$.

Note that in the master equation Eq. (59), terms $\mu=\pm$ and $\mu=\phi$ show the relaxation, excitation and pure dephasing of the Floquet qubit, respectively. 

As an example to clarify the Floquet master equation, we consider two major noise sources that influence the fluxonium coherence time: $1/f$ flux noise and dielectric loss 
\cite{Nguyen,Zhang2,Hazard}. Therefore, the interaction Hamiltonian can be written as $H_{\Int}=(\eta_f+\eta_d)\sigma_z$, where $\eta_f$ and $\eta_d$ are the environmental operators corresponding to $1/f$ flux noise and dielectric loss, respectively. We define the noise spectrum as

\begin{subequations}
\begin{align}
&S_f(\omega)={\mathcal A}^2_f \vert \omega/2\pi\vert^{-1} \\
&S_d(\omega)=\alpha(\omega ,T) {\mathcal A}_d (\omega/2\pi)^2,
\end{align}
\end{subequations}
where $\alpha (\omega , T)=\vert \coth (\omega / 2k_BT)+1\vert /2$ is a thermal factor, $k_B$ and $T$ represent the Boltzman constant and temperature and ${\mathcal A}_f $ and ${\mathcal A}_d $ are the noise amplitude \cite{Huang}.

To calculate the decoherence rate $\gamma_\mu$, we assume $F_\mu(\omega , t)$ are peaked at the filter frequencies $\omega _{k\mu}$ that the peak width is $2\pi t^{-1}$. We consider the case in which the spectrum $S(\omega)$ is sufficiently flat within each peak-width frequency range. Thus, we have $(t/\pi)\sinc (\omega t)\approx \delta (\omega)$ in Eq. (60). Consequently, a Markovian Floquet master equation can be obtained as 

\begin{align}
\frac{\D}{\D t}\rho_{\rm q}(t) =\sum_{k\in \mathbb{Z},\mu=\pm,\phi} \vert g_{k\mu}\vert ^2 S(\omega_{k\mu}) {\mathcal D}[c_\mu] \rho_{\rm q}(t),
\end{align}
where the decoherence rates are given by \cite{Huang}

\begin{subequations}
\begin{align}
&\gamma_{\pm}=\sum_k  \vert g_{k\mu}\vert ^2 S(\omega_{k\pm}) \\
&\gamma_{\phi}=\sum_k  2 \vert g_{k\phi}\vert ^2 S(\omega_{k\phi}),
\end{align}
\end{subequations}

We can also solve the Floquet master equation in Eq. (59) numerically by QuTiP. It needs to define the operators through which the qubit couples to the environment and the noise spectral density function of the environment \cite{Nation}.

As we mentioned before, the Floquet theory is a non-perturbative approach to describe the effect of an oscillating field on the superconducting qubits. In the next section, we use a perturbative method to analyze the dynamics of the superconducting qubits as open quantum systems.

\subsection{Perturbation Theory}

In quantum mechanics, perturbation theory is a method to analyze the behavior of a complex quantum system by comparing it to a simpler one. The idea is to start with a simple system where we already have an exact solution of the  Schrödinger equation for it, and then introduce a perturbing Hamiltonian representing a weak perturbation to the system. If the perturbation is not very big, we can express the different physical values of the perturbed system such as energy levels and eigenstates as corrections to those of the simple system. These small corrections can be calculated using approximate approaches such as asymptotic series. Perturbation theory is a useful tool for explaining real quantum systems. It helps when we can't find exact answers to the Schrödinger equation for more complicated Hamiltonians \cite{Messiah,Sakurai}.

In recent years, perturbation theory has been used to analyze the dynamics of the superconducting qubits as an open quantum system. In this regard, we describe two important perturbative methods that are applicable in the CQED and the calculation of the decoherence rates of the qubit in the presence of noises, respectively.

\subsubsection{Schrieffer-Wolff perturbation theory}

The exact diagonalization of the interaction Hamiltonian of two coupled systems is so complex. In general, a unitary transformation is employed to diagonalize the Hamiltonian. In this context, the Schrieffer-Wolff transformation which is a general perturbative approach has been developed to find the low-energy effective Hamiltonian from the exact Hamiltonian by a unitary transformation decoupling \cite{Schrieffer, Bravyi}

To describe the Schrieffer-Wolff approach in CQED, let us consider a generic Hamiltonian as \cite{Blais}

\begin{align}
H=H_0+V,
\end{align}
with $H_0$ and $V$ denote a free Hamiltonian (unperturbed) and a perturbation, respectively. Note that for the Schrieffer-Wolff transformation to be valid, it is assumed that $V$ is a small perturbation. The total Hilbert space that describes the system can be split into subspaces in which $H_0$ and $V$ does not and does couple states in different subspaces, respectively. The Schrieffer-Wolff method aims to derive an effective Hamiltonian for which the different subspaces are completely decoupled. 
 
It is shown that the different subspaces with a subscript $\mu$, can be defined by projection operators as \cite{Tannoudji,Zhu}

\begin{align}
P_\mu=\sum_n \vert \mu,n\rangle \langle \mu,n\vert,
\end{align}
in which $\vert \mu,n\rangle, n=0,1,...$ represents an orthonormal basis for the subspace $\mu$. The unitary transformation that is often used in CQED is as follows

\begin{align}
H_U=U^{\dagger} H U-i U^{\dagger} U,
\end{align}
where $U$ is a time-dependent unitary transformation and we have $\vert \psi_U\rangle =U^{\dagger}\vert \psi\rangle $.

The Schrieffer-Wolff method is based on finding a unitary transformation $U=e^{-S}$ that approximately decouples the various subspaces by using the Baker-Campbell-Hausdorff lemma. For any two possibly non-commutative operators $H$ and $S$, the Baker-Campbell-Hausdorff lemma can be written as \cite{Blais}

\begin{align}
e^S He^{-S}&=H+[S,H]+\frac{1}{2!} [S,[S,H]]+...\nonumber\\
&=\sum_{n=0}^\infty \frac{1}{n!} C^n_S[H],
\end{align}
where 

\begin{align}
C^n_S[H]=[S,[S,[S,...,H]]],
\end{align}
for $n$ times and $C^0_S[H]=H$ \cite{Boissonneault}.

We expand $H$ and $S$ in Maclaurin power series as

\begin{subequations}
\begin{align}
&H=H^{(0)}+\gamma H^{(1)}+\gamma^2 H^{(2)}+... \\
&S=\gamma S^{(1)}+\gamma ^2 S^{(2)}+...,
\end{align}
\end{subequations}
where $\gamma$ is a dimensionless parameter to simplify order counting that can take on values ranging continuously from 0 (no perturbation) to 1 (the full perturbation). We can ultimately set  $\gamma\rightarrow 1$. 

Considering Eqs. (68)-(72), collecting terms at each power of $\gamma^k$ and iteratively solving for $S^{(k)}$ and $H^{(k)}$, we find the Schrieffer-Wolff transformation. We have used two important requirements: first, the resulting Hamiltonian $H_U$ is block-diagonal means that it does not couple different subspaces $\mu$ at each order $k$. Second, $S$ is itself block-off-diagonal \cite{Bravyi}. Finally, the explicit results for the second-order perturbation theory ($k=2$) with $\gamma=1$ and for $\nu\neq \mu$ can be obtained as \cite{Blais}

\begin{align}
\langle \mu,n \vert S^{(1)}\vert \nu ,l\rangle=\frac{\langle \mu,n \vert V\vert \nu ,l\rangle}{E_{\mu ,n}-E_{\nu ,l}} ,
\end{align}

\begin{align}
\langle \mu,n \vert S^{(2)}\vert \nu ,l \rangle=\sum_k \Big (\frac{\langle \mu,n \vert V \vert \nu ,k\rangle}{E_{\mu ,n}-E_{\nu ,l}}   \frac{\langle \nu,k \vert V \vert \nu ,l \rangle}{E_{\mu ,n}-E_{\nu ,k}} - \frac{\langle \mu,n \vert V \vert \mu ,k\rangle}{E_{\mu ,n}-E_{\nu ,l}}    \frac{\langle \mu,k \vert V \vert \nu ,l\rangle}{E_{\mu ,k}-E_{\nu ,l}}    \Big),
\end{align}
while for $\nu= \mu$, the block-diagonal matrix element vanish. Moreover, we have

\begin{subequations}
\begin{align}
&H^{(0)}=H_0 \\
&H^{(1)}=\sum_\mu P_\mu VP_\mu \\
\langle \mu,n \vert H^{(2)}\vert \mu,m\rangle &=\sum_{\nu \neq \mu,l} \langle\mu,n\vert V\vert\nu,l \rangle \langle\nu,l\vert V\vert \mu,m\rangle \times \frac{1}{2} \Big( \frac{1}{E_{\mu ,n}-E_{\nu ,l}} +\frac{1}{E_{\mu ,m}-E_{\nu ,l}}\Big),
\end{align}
\end{subequations}
for the transformed Hamiltonian. For $\mu\neq \nu$, block-off-diagonal matrix elements vanish, $\langle \mu,n \vert H^{(2)}\vert \nu,m\rangle=0$. In addition, $E_{\mu,n}$ represents the bare energy of $\vert\mu,n\rangle$ under $H_0$ \cite{Blais}.

Now we use the Schrieffer-Wolff method to study a situation that often happens in CQED.

Let us consider an artificial atom coupled to a single-mode oscillator in the dispersive regime in which the qubit-resonator detuning is large with respect to the coupling strength. Since in a dispersive regime, the qubit and resonator are weakly entangled, we can use the second-order perturbation theory as an excellent approximation. The transmon artificial atom and the two-level system can be good examples of this general case. 

We can define the total Hamiltonian as \cite{Blais}

\begin{align}
H=\omega_r a^{\dagger}a+\sum_j  \omega_j \vert j\rangle\langle j\vert+(Ba^{\dagger}+B^{\dagger} a),
\end{align}
where the Hamiltonian of the artificial atom which is considered a generic multilevel system can be defined as $H_a=\sum_j \omega_j \vert j\rangle\langle j\vert$ in which $\vert j \rangle$ is atomic eigenstate. Furtheremore, $B$ denotes an operator of the atom that interacts with the oscillator and $\omega_r$ is the resonator frequency.

It is shown that using the identity $I=\sum_j \vert j\rangle\langle j\vert$, the interaction term can be written as  \cite{Koch}

\begin{align}
H=\omega_r a^{\dagger}a+\sum_j \omega_j \vert j\rangle\langle j\vert +\sum_{ij}  \big( g_{ij}\vert i \rangle \langle j \vert a^{\dagger}+g^*_{ij} \vert j \rangle \langle i \vert a \big),
\end{align}
where $ g_{ij}=\langle i \vert B \vert j \rangle$  with $g_{ij}=g_{ji}$, if $ B=B^{\dagger}$. Comparing Eq. (77) with Eq. (67), we consider the two first terms and the last term of Eq. (77) as $H_0$ and $V$, respectively. In this situation, the subspaces $\mu$ are one dimensional $(P_\mu=\vert \mu \rangle \langle \mu \vert)$, with $\vert \mu \rangle=\vert n, j \rangle=\vert n \rangle \otimes \vert j \rangle$, where $\vert n \rangle $ is an oscillator number state.

After some calculations, one can show that \cite{Koch,Zhu}

\begin{align}
H_{\disp}= e^S H e^{-S}\simeq  \omega_r a^{\dagger} a+\sum_j (\omega_j+\Lambda _j) \vert j \rangle \langle j \vert +\sum_j \chi_j a^{\dagger}a \vert j \rangle \langle j \vert,
\end{align}
in which

\begin{subequations}
\begin{align}
&\Lambda_j=\sum_j \chi_{ij} \\
&\chi_j=\sum_i (\chi_{ij}-\chi_{ji}),
\end{align}
\end{subequations}
and

\begin{align}
\chi_{ij}=\frac{\vert g_{ji} \vert ^2}{\omega_j-\omega_i-\omega_r},
\end{align}

Considering the two atomic levels $j=0,1$ and $\sigma_z=\vert 1 \rangle \langle 1\vert -\vert 0 \rangle\langle 0\vert$, we have

\begin{align}
H_{\disp}\simeq \omega'_r a^{\dagger} a+\frac{\omega'_{\rm q}}{2}\sigma_z + \chi a^{\dagger} a \sigma_z,
\end{align}
where $\omega'_r=\omega_r+(\chi_0+\chi_1)/2$, $\omega'_{\rm q}=\omega_1-\omega_0+\Lambda_1-\Lambda_0$ and $\chi=(\chi_1-\chi_0)/2$ \cite{Blais}. In general, by decoupling the low-energy and high-energy subspaces, we have obtained the low-energy effective Hamiltonian in Eq. (81).

\subsubsection{Calculation of decoherence rate}

Another example of the application of perturbation theory in superconducting qubits is the calculation of the decoherence rate for the qubits in the presence of noise. To describe the details of this approach, we consider the dynamics of a qubit from a general point of view in a free evolution regime.

After the preparation of qubit states, the effective spin evolves in an effective magnetic field under the influence of the static field $H_0$, which is adjusted by the control parameters $\lambda_0$ such as the gate charge on the island between the two small Josephson junctions and the phase. Moreover, its classical and quantum fluctuations are set by  $\delta \lambda$ \cite{Ithier}. 

As discussed in section III.C, the Bloch-Redfield theory \cite{Wangsness,Abragam,Geva} describes the dynamics of a qubit in the presence of noise in terms of two times: the relaxation time $T_1=1/\Gamma_1$ dominated at low temperatures for the decay of the diagonal component of the spin density matrix and the qubit coherence time $T_2=1/\Gamma_2$ for the decay time of the off-diagonal part which is inferred from the Ramsey experiment \cite{Slichter}. The relation between these parameters is given in Eq. (5) as $\Gamma_2=\Gamma_\phi+\frac{\Gamma_1}{2}$ in which $\Gamma_\phi$ is the dephasing rate. 

The Hamiltonian of the qubit such as quantronium $H_{\rm q}$ is that of a spin-$1 /2$ particle under the influence of the static field   $H_0= \omega_{\rm q} \hat{z}$ can be written as 
 
\begin{align}
 H_{\rm q}=-\frac{1}{2} H_0 \sigma,
\end{align}
where $\omega_{\rm q}$ is the qubit frequency and $\sigma=\sigma _x \hat{x}+\sigma _y \hat{y}+\sigma _z \hat{z}$.

Let us expand $H_{\rm q}$ to second-order in the perturbation theory as \cite{Ithier}

\begin{align}
 H_{\rm q}=-\frac{1}{2} \Big(H_0(\lambda_0)+\frac{\partial H_0}{\partial \lambda}\delta \lambda+ \frac{\partial^2 H_0}{\partial \lambda ^2}\frac{\delta \lambda^2}{2}  +...\Big)\sigma ,
\end{align}
We define the notations $D_\lambda\equiv  \partial H_0 / \partial \lambda $ and $D_{\lambda 2}\equiv  \partial ^2 H_0 / \partial \lambda ^2$. Therefore, in the eigenbasis of $H_0(\lambda_0) \sigma$, one can show

\begin{align}
 H_{\rm q}=-\frac{1}{2}  (\omega _{\rm q} \sigma_z+\delta \omega_z \sigma _z +\delta \omega_ {\perp} \sigma_{\perp}),
\end{align}
where 

\begin{subequations}
\begin{align}
&\omega_{\rm q}\equiv \vert H_0(\lambda _0)\vert  \\
&\delta \omega _z\equiv D_{\lambda , z}\delta _{\lambda}+D_{\lambda 2, z} \delta \lambda ^2/2+... \\
&\delta \omega_{\perp}\equiv D_{\lambda , \perp} \delta \lambda+...,
\end{align}
\end{subequations}
and $\sigma_{\perp}$ represents the transverse spin components. Furthermore, we define the coefficients $D$ as 

\begin{align}
D_{\lambda , z}=\frac{\partial \omega_{\rm q}}{\partial \lambda},
\end{align}
and

\begin{align}
D_{\lambda 2 , z}=\frac{\partial ^2 \omega_{\rm q}}{\partial \lambda ^2}-\frac{D^2_{\lambda , \perp}}{\omega_{\rm q}},
\end{align}

According to Bloch-Redfield theory, the pure dephasing $\Gamma_\phi$ is usually associated with the broadening in a set of spins and the longitudinal low-frequency noise for a single spin. Therefore, $\partial \omega_{\rm q}/ \partial \lambda$ and $\partial ^2 \omega_{\rm q}/\partial \lambda ^2$ are sufficient to analyze the low-frequency noise whereas the relaxation rate includes $D_{\lambda , \perp}$. Note that the Bloch-Redfield approach applies only if the noise is weak and short-correlated such as white noise \cite{Slichter}.

For the relaxation rates, it is shown that \cite{Ithier}

\begin{align}
\Gamma_1=\pi Ss_{\delta \omega_{\perp}}(\omega_{\rm q})=\pi D^2_{\lambda , \perp} Ss_{\lambda} (\omega_{\rm q}),
\end{align}
where $S_{\lambda}$ is the correlator. This approximation is independent of the statistics of the fluctuations and is sufficient for weak noise at low temperatures in which $k_BT \ll  \omega_{\rm q}$. For the dephasing rate $\Gamma_\phi$, we consider a noise source coupled linearly to the qubit in which $\partial \omega_{\rm q}/ \partial \lambda=D_{\lambda , z}\neq 0$. So, the Bloch-Redfield theory gives \cite{Ithier}

\begin{align}
\Gamma_\phi=\pi S_{\delta \omega _z}(\omega =0)&=\pi D^2_{\lambda , z} S_{\lambda} (\omega=0)\nonumber\\
&=\pi D^2_{\lambda , z} Sc_{\lambda}(\omega=0),
\end{align}
This relation similar to Eq. (88) is a golden rule type and holds, if the noise power $Sc_{\lambda}$ is regular near $\omega \approx0$ up to frequencies of order $\Gamma_\phi$.

So far, we haven't investigated the interaction between a superconducting qubit and an electromagnetic field through the transmission line. In the next section, we use input-output theory to obtain information about the fields radiated by the superconducting qubits that is of practical importance in qubit readout.

\subsection{Input-Output Theory}

As we have mentioned, the master equation is an important and common method to analyze the dynamics of a quantum system interacting with an uncontrolled environment. However, it does not tell us any information about the fields radiated by the system that is of practical importance. In this regard, the input-output theory is developed to describe the interaction between the system and a controlled environment such as traveling electromagnetic fields through transmission lines in superconducting qubits to simulate a real experiment \cite{Collett,Gardiner1,Holland,Gardiner2,Combes,Jacobs}. In this formalism, the {\it input} denotes the field that drives the system, and {\it output} denotes the radiated field by the system including the input fields and a contribution from the systems that are measured experimentally \cite{Kwon}. Although the input-output theory was originally developed for the damping behavior of a mode of an optical cavity, it is currently an appropriate tool to analyze dissipative quantum systems that are weakly coupled to an environment. In this regard, the input-output theory has attracted much attention to describing the behavior of superconducting \cite{Kamal,Naik,Clark2,Fitzpatrick,Kelly,Cook} and photonic circuits \cite{Yoshie,Reithmaier,Fischer,Schroder}. In general, this method analyzes the interaction of circuit components such as localized quantum systems with the environment which was replaced by transmission lines and waveguides. In other words, the environment is modeled as a quantum field and we consider the system as an elastic scatterer of fields \cite{Vool,Jacobs, Cook}.

Although the Markovian master equations can be obtained from input-output theory, this theory has important and unique features such as \cite{Jacobs}

\subparagraph{i.} avoiding the approximations used in deriving the master equations;

\subparagraph {ii.} the physical relationship between the environment and continuous measurements; 

\subparagraph{iii.} calculation of the output from the system into the environment;

\subparagraph {iv.} to link systems in a network by connecting the outputs of some systems to the inputs of others; 

\subparagraph{v.} applicable in a non-Markovian regime using the quantum Langevin approach of input-output theory \cite{Zhang3,Giovannetti}.

Moreover, the unique feature of the input-output theory in comparison to some other methods to analyze the dissipative quantum system is keeping the environment's degrees of freedom instead of tracing them out. This feature is crucial when we want to obtain information from the output field while measuring the system that interacted with the cavity \cite{Clerk1}.

Despite all the advantages, this method has an important deficiency: it generally cannot describe conditions where fields can pass through loops created unintentionally by retro-reflections from circuit components. This shortcoming is caused by the fact that loops let individual fields interact with the same circuit components an infinite number of times. This creates a non-Markovian regime in which states of the system at $t$ affect the system at $t'$ through the fields \cite{Cook,Grimsmo,Pichler}.

\subsubsection{Input-output theory in CQED}

In this subsection, we describe the formalism of input-output theory which is a useful method for understanding qubit readout in CQED regarding \cite{Blais,Yurke1,Yurke2}.

\subsubsection*{i. Quantum system-transmission lines coupling}

Analysis of the interaction between the quantum electrical circuits and their environments such as the measurement apparatus and control circuitry is needed to analyze the CQED. Here, we consider the transmission lines coupled to quantum systems, as an important model for dissipation that is used in the control and measurement of superconducting qubits. In particular, we consider a semi-infinite transmission line coupled to an oscillator, as illustrated in FIG. 7. The semi-infinite transmission line is a limit of the coplanar waveguide resonator with a finite length in which one of the boundaries has approached infinity. This results in a dense frequency spectrum that should be considered as a continuum in its infinite limit \cite{Blais}.

Let us first define the Hamiltonian of the transmission line as

\begin{align}
H_{\rm tml}=\int_0^{\infty} \D \omega  b^{\dagger}_{\omega} b_{\omega} \omega,
\end{align}
where the commutation relation is $[b_{\omega}, b^{\dagger}_{\omega'}]=\delta (\omega -\omega')$ for the mode operators. Moreover, the flux and charge operators of the transmission line in the continuum limit can be defined as \cite{Yurke2}

\begin{subequations}
\begin{align}
&\Phi_{\rm tml}(x)=\int_0^{\infty} \D \omega \sqrt{\frac{1}{\pi \omega c v}} \cos \Big(\frac{\omega x}{v}\Big) (b^{\dagger}_{\omega}+ b_{\omega}) \\
&Q_{\rm tml}(x)=i \int_0^{\infty} \D \omega \sqrt{\frac{\omega c}{\pi  v}} \cos \Big(\frac{\omega x}{v}\Big) (b^{\dagger}_{\omega}- b_{\omega}),
\end{align}
\end{subequations}
where are the canonical fields of the transmission line and we have in the Heisenberg picture

\begin{align}
Q_{\rm tml}(x,t)=c \dot{\Phi}_{\rm tml}(x,t),
\end{align}
and they satisfy the canonical commutation relation $[\Phi_{\rm tml}(x),Q_{\rm tml}(x')]=i \delta (x-x') $. In addition, $v=1/\sqrt{lc}$ is the speed of light in the transmission line in which $c$ and $l$ show the capacitance and inductance per unit length, respectively.

The total Hamiltonian for the transmission line-oscillator coupling at $x=0$ can be written as \cite{Blais}

\begin{align}
H=H_{\rm s}+H_{\rm tml}-\int_0^{\infty}\D \omega \lambda (\omega) (b^{\dagger}_{\omega}- b_{\omega}) (a^{\dagger}- a),
\end{align}
where the Hamiltonian of the oscillator is $H_{\rm s}=\omega_r a^{\dagger} a$. Furthermore, the coupling strength is defined as $\lambda(\omega)=\big(C_k / \sqrt{cC_r}\big) \sqrt{\omega_r \omega / 2 \pi v}$ where $C_k$ and $C_r$ are the coupling capacitance and the resonator capacitance between the oscillator and the line, respectively.

To consider the interaction as a perturbation, we assume $\lambda(\omega) \ll \omega_r$. Therefore, the Q factor of the system is large and the oscillator responds in a small bandwidth only around $\omega_r$. So, we take $\lambda(\omega)\simeq \lambda (\omega_r)$ in Eq. (93). Consequently, regarding RWA, we obtain \cite{Gardiner}

\begin{align}
H\simeq H_{\rm s}+H_{\rm tml}+ \int_0^{\infty}\D \omega \lambda (\omega_r) (a b^{\dagger}_{\omega}+a^{\dagger} b_{\omega}).
\end{align}

\begin{figure}
\centering
\includegraphics[scale=0.4]{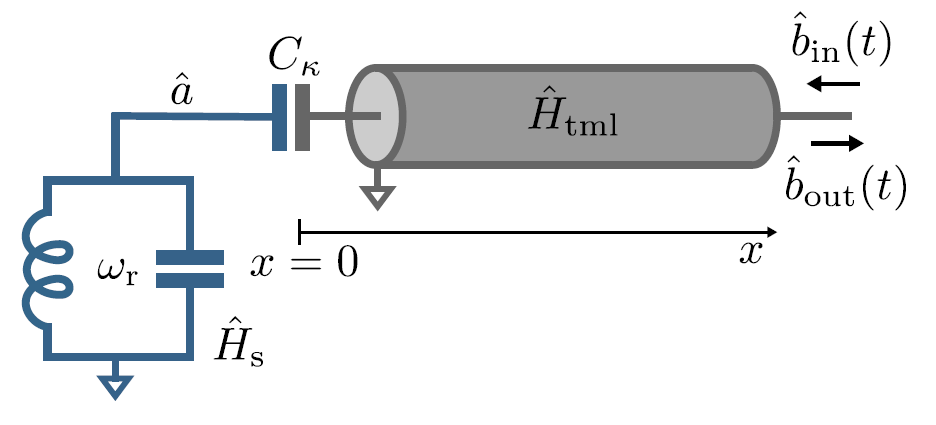}
\caption{A schematic of an LC circuit coupled to a semi-infinite transmission line to analyze the damping and driving of the system. $b_{\rm in}(t)$ and $b_{\rm out}(t)$ are the oscillator’s input and output fields, respectively. Source \cite{Blais}.}
\end{figure}

\subsubsection*{ii. The formalism of input-output theory}

There are two standard approaches in the input-output theory. In the first approach, regarding Eq. (94), we consider Heisenberg equations of motion for the annihilation operators of the system $a$ and the field $b_\omega$, which is a well-known approach in quantum optics \cite{Collett, Gardiner1}. In the second approach, the transmission line modes are decomposed into two parts of the moving fields that are connected by a boundary condition at the position of the oscillator at $x=0$ and the transmission line at $x>0$ \cite{Yurke1}. The crucial feature of this method is that the input and output fields of the oscillator can be defined in two terms of the radiation field components propagating along the transmission line. Here, we describe the details of this approach.

Let us consider the coupling between an oscillator at $x=0$ and a semi-infinite transmission line extending from $x=0$ to $x=\infty$, as shown in FIG. 7. Considering the flux and charge operators, the Hamiltonian of the transmission line is \cite{Blais}

\begin{align}
H_{\rm tml}=\int_{\infty}^{\infty} \D x \theta (x) \Big\lbrace \frac{Q_{\rm tml}(x)^2}{2c}  +  \frac{[\partial_x \Phi_{\rm tml} (x)]^2} {2l}\Big\rbrace,
\end{align}
where $\theta(x)$ is the Heaviside step function. Moreover, Hamiltonian of the oscillator with $\omega_r=1/\sqrt{L_rC_r}$ is

\begin{align}
H_{\rm s}=\frac{Q^2_r}{2C_r}+\frac{\Phi^2_r}{2L_r},
\end{align}
The interaction Hamiltonian can be written as

\begin{align}
H_{\rm int}=\int_{\infty}^{\infty} \D x \delta(x) \frac{C_k}{cC_r}Q_r Q_{\rm tml}(x),
\end{align}
Hamilton’s equations for the field in the transmission line are obtained as

\begin{align}
\dot{\Phi}_{\rm tml}(x)=\theta(x) \frac{Q_{\rm tml}(x)}{c}+\delta(x) \frac{C_k}{cC_r}Q_r,
\end{align}

\begin{align}
\dot{Q}_{\rm tml}(x)=\partial_x \Big[\theta(x)\frac{\partial_x \Phi_{\rm tml}(x)}{l} \Big],
\end{align}
For $x>0$, we combine Eqs. (98) and (99) to a wave equation for $\Phi_{\rm tml}$ as

\begin{align}
\ddot{\Phi}_{\rm tml}(x)=v^2 \partial^2_x \Phi_{\rm tml}(x),
\end{align}
Moreover, at the location $x=0$ of the oscillator, we obtain

\begin{align}
\ddot{\Phi}_{\rm tml}(x)=\theta(x) v^2 \big[ \delta(x) \partial_x \Phi_{\rm tml}(x)+\partial^2_x \Phi_{\rm tml}(x)\big] +\delta(x) \frac{C_k}{cC_r}\dot{Q}_r,
\end{align}
where $\partial_x\theta(x)=\delta(x)$. To find the boundary condition, we integrate Eq. (101) at the range $-\varepsilon < x < \varepsilon$ and then we use the limit $\varepsilon\rightarrow 0$. So, we find

\begin{align}
v^2 \partial_x \Phi_{\rm tml}(x=0)=-\frac{C_k}{c C_r}\dot{Q}_r.
\end{align}
Regarding Eqs. (92) and (100), we obtain the general forms of the flux and charge for $x >0$ as

\begin{subequations}
\begin{align}
\Phi_{\rm tml}(x,t)=\Phi_{\rm L}(x,t)+\Phi_{\rm R}(x,t)  \\
Q_{\rm tml}(x,t)=Q_{\rm L}(x,t)+Q_{\rm R}(x,t),
\end{align}
\end{subequations}
where the subscripts L and R represent left- and right-moving fields. We rewrite Eq. (91) as

\begin{subequations}
\begin{align}
\Phi_{\rm L/R}(x,t)=\int_0^{\infty} \D \omega \sqrt{\frac{1}{4\pi \omega cv}}e^{\pm i\omega x/v+i\omega t}b^{\dagger}_{\rm L/{R{\omega}}}+{\rm H.C.}   \\
Q_{\rm L/R}(x,t)=i \int_0^{\infty} \D \omega \sqrt{\frac{\omega c}{4\pi v}}e^{\pm i\omega x/v+i\omega t}b^{\dagger}_{\rm L/{R{\omega}}}-{\rm H.C.},
\end{align}
\end{subequations}
where H.C denotes Hermite Conjugate and we have $[b_{\nu \omega},b^{\dagger}_{\mu \omega'}]=\delta_{\nu \mu}\delta (\omega-\omega')$ for $\nu={\rm R,L}$.

The left- and right-moving fields are dependent due to the boundary condition at $x=0$. To clarify this, let us consider the form of $\Phi_{\rm tml}(x,t)$ as 

\begin{align}
Z_{\rm tml}\frac{\partial_x \Phi_{\rm tml}(x,t)}{l}=\dot{\Phi}_{\rm L}(x,t)-\dot{\Phi}_{\rm R}(x,t),
\end{align}
where $Z_{\rm tml}=\sqrt{l/c}$ is the characteristic impedance of the transmission line. Defining the current $I(x)=\partial_x \Phi_{\rm tml}(x)$ and voltage $V_{\rm L/R}(x)=\dot{\Phi}_{\rm L/R}(x)$, we can consider Eq. (105) as Ohm's law. Using Eq. (102), we obtain the boundary condition at $x=0$ as

\begin{align}
V_{\rm out}(t)-V_{\rm in}(t)=Z_{\rm tml} I(t) ,
\end{align}
in which $I(t)=(C_k/C_r) \dot{Q}_r(t)$ and $V(t)=V(x=0,t)=V_{\rm in}(t)+V_{\rm out}(t)$ and we have used the standard notation $V_{\rm in/out}(t)=V_{L/R}(x=0,t)$. Furthermore, the oscillator's charge and flux operators in terms of the ladder operator can be written as \cite{Blais}

\begin{subequations}
\begin{align}
\Phi=\Phi_{\rm zpf} (a^{\dagger}+a) \\
Q=i Q_{\rm zpf} (a^{\dagger}-a),
\end{align}
\end{subequations}
where $\Phi_{\rm zpf}=\sqrt{1 /2\omega_r C}=\sqrt{ Z_r/2}$ and $Q_{\rm zpf}=\sqrt{ \omega_r C /2}=\sqrt{1 /2 Z_r}$ are the characteristic magnitude of the zero-point
fluctuations of the flux and the charge, respectively. Using Eqs. (104) and (107), we obtain 

\begin{align}
-i&\int_0^{\infty} \D \omega \sqrt{\frac{\omega}{4\pi c v}}e^{-i (\omega-\omega_r)t} (b_{ \rm R\omega}-b_{\rm L\omega})=-\omega_r Z_{\rm tml} \frac{C_k}{C_r}\sqrt{\frac{\omega_r C_r}{2}}a.
\end{align}
Consequently, after some rearrangement, we obtain the standard input-output relation \cite{Collett,Gardiner1}

\begin{align}
b_{\rm out}(t)-b_{\rm in}(t)=\sqrt{\kappa} a(t),
\end{align}
where the input and output fields defined as

\begin{subequations}
\begin{align}
b_{\rm in}(t)=\frac{i}{\sqrt{2\pi}}\int_{-\infty}^{\infty} \D \omega b_{\rm L\omega}e^{-i(\omega-\omega_r)t}  \\
b_{\rm out}(t)=\frac{i}{\sqrt{2\pi}}\int_{-\infty}^{\infty} \D \omega b_{\rm R\omega}e^{-i(\omega-\omega_r)t},
\end{align}
\end{subequations}
which the commutation relations is satisfied as $[b_{\rm in}(t),b^{\dagger}_{\rm in}(t')]=[b_{\rm out}(t),b^{\dagger}_{\rm out}(t')]=\delta(t-t')$. The photon-loss rate $\kappa$ is

\begin{align}
\kappa=\frac{Z_{\rm tml}C^2_k \omega^2_r}{C_r},
\end{align}
Note that to derive Eq. (109), we have assumed terms with $\omega \simeq \omega_r$ contribute significantly to Eq. (108). Therefore, terms rotating at $\omega+\omega_r$  have been dropped. Based on this approximation, we can extend the limit of integration from $[0,\infty)$ to $(-\infty,\infty)$.

Moreover, Eq. (105) can be rewritten as

\begin{align}
\partial_x \Phi_{\rm tml}(x,t)&=Z_{\rm tml}\big[Q_{\rm L}(x,t)-Q_{\rm R}(x,t)\big] \nonumber\\
&=Z_{\rm tml}\big[2Q_{\rm L}(x,t)-Q_{\rm tml}(x,t)\big],
\end{align}
in which we have used Eq. (103). At $t=0$, we have 

\begin{align}
Q_{\rm tml}(x=0,t)=2Q_{\rm L}(x=0,t)+\frac{1}{v} \frac{C_k}{C_r}Q_r(t),
\end{align}
For the oscillator, the Heisenberg euqations of motion can be obtained as

\begin{align}
\dot{\Phi}_r=\frac{Q_r}{C_r}+\frac{C_k}{cC_r}Q_{\rm tml}(x=0),
\end{align}

\begin{align}
\dot{Q}_r=-\frac{\Phi_r}{L_r},
\end{align}
In a single equation of motion, we have

\begin{align}
\ddot{Q}_r=-\omega^2_r\Big[Q_r+\frac{C_k}{c}\Big(\frac{1}{v}\frac{C_k}{C_r}Q_r+2Q_{\rm in}\Big)\Big],
\end{align}
Considering Eq. (107), Eq. (116) can be rewritten as the Langevin equation for the resonator field $a(t)$  \cite{Yurke2}

\begin{align}
\dot{a}(t)=i[H_{\rm s},a(t)]-\frac{\kappa}{2}a(t)+\sqrt{\kappa} b_{\rm in},
\end{align}
where is a Heisenberg picture analog to the Markovian master equation. This relation represents that the dynamics of the resonator is described by the input field such as noise or drive, while Eq. (109) represents how the output can be obtained from the input and the dynamics of the system. In other words, the output field which can be measured experimentally, includes information about the response of the system to the input field and indirectly gives us information about the dynamics of the system. The measurement process is done by measuring the voltage at $x>0$. This voltage can be defined as \cite{Blais}

\begin{align}
V(x,t) \simeq \sqrt{\frac{\omega_r Z_{\rm tml}}{2}}\Big[ e^{i\omega_r x/v-i \omega_rt} b_{\rm out}(t)  + e^{-i\omega_r x/v-i \omega_rt} b_{\rm in}(t)+{\rm H.C.} \Big].
\end{align}
To derive this relation for the voltage, we have neglected non-Markovian time-delay effects \cite{Blais}.

For instance, input-output theory is particularly relevant for modeling the readout of transmon or fluxonium qubits coupled to microwave resonators, as commonly implemented in CQED architectures.

In summary, we have considered an oscillator coupled to a semi-infinite transmission line. In a general case, quantum systems can be coupled to multiple transmission lines to form quantum networks. This complicated case can be analyzed using the SLH formalism. For more details, please see \cite{Combes}.

In the next section, we use a stochastic approach to describe the average behavior of a superconducting qubit in the presence of a random interaction with the environment.

\subsection{Stochastic Approach}

In physics, stochasticity appears whenever we consider the dynamics of a system interacting with an environment in which the environment is considered to be in thermal equilibrium. So, any fluctuation in the environment leads to a random influence on the system. A stochastic process is a process in which a random variable $X(t)$ whose statistical properties change over time and the property of the system is determined by this variable. The main idea of the stochastic process is the generalization of the deterministic time evolution. However, in the stochastic process, an indeterministic time evolution of some variables can be used instead of a deterministic one. In other words, a stochastic process is  a collection of $X(t)$ on a probability space depending on a parameter $t \in T$. In the most important examples of stochastic process, $t$ plays the role of the time variable, and $T$ is usually an interval of the real time \cite{Breuer,Biele}.

In recent years, stochastic approaches have been much attention in various fields such as thermal relaxation, Brownian motion, thermal and electrical noise, and phase transitions \cite{Biele,Trapani,Diosi,Stockburger1,Strunz,Stockburger2}. In many cases, the coupling between the system and its environment has a random nature. Therefore, we can use stochastic approaches will give average behavior to analyze the dynamics of an open quantum system. In this regard, the dynamics of such systems can be analyzed by some stochastic approaches such as the stochastic wave function methods in which an ensemble of pure states is propagated using a stochastic time evolution instead of solving the quantum master equation \cite{Breuer,Breuer2,Breuer3,Percival}. Moreover, the integro-differential stochastic equations have been developed in which the effects of the environment are described by a stochastic and time non-local mean-field \cite{Lacroix}. In quantum optics, the stochastic methods can be used to describe the effects of noisy classical environments on quantum systems such as an atom
 \cite{Trapani}.

In the following subsections, we describe the details of the important stochastic approaches that can be used to describe the dynamics of superconducting qubits as open quantum systems.

\subsubsection{Stochastic Schrödinger equation (SSE)}

Similar to classical stochastic approaches such as the Fokker–Planck and Langevin equations, the stochastic Schrödinger equation (SSE) is developed to analyze the dynamics of an open quantum system effectively by simulating the average behavior of the system. In this framework, one can inspire the situation of the Langevin equation of Brownian motion and define a stochastic term to describe the behavior of the system \cite{Gisin,Struntz1,Struntz2,Vega}. The SSE aims to reproduce the dynamics from the density matrix after taking an average. Here, we formulate the general expression of SSE which is given in \cite{Biele,Gaspard} based on the equation of motion for the system, the environment, their coupling and the Feshbach projection-operator method.

Let us consider $H_{\rm s}$, $H_\varepsilon$ and $\lambda W$ as the Hamiltonians of the system, Hamiltonian of the environment and the interaction potential ($\lambda$ is coupling parameter), respectively. The Schrödinger equation for the total system can be written as

\begin{align}
i \D \vert \Psi(t)\rangle =\Big(H_{\rm s}+H_\varepsilon+\lambda W\Big) \vert \Psi(t)\rangle \D t.
\end{align}
We define a complete and orthonormal environmental basis as

\begin{subequations}
\begin{align}
&\mathbf{1}_\varepsilon=\sum_n \vert n \rangle \langle n \vert    \\
&H_\varepsilon \vert n \rangle = \epsilon _n \vert n \rangle   \\
&\langle m \vert n \rangle =\delta_{mn},
\end{align}
\end{subequations}
and the total wave function can be expanded as 

\begin{align}
\vert \Psi(t) \rangle=\sum_n \vert \phi^n(t)\rangle \otimes \vert n \rangle,
\end{align}
The normalization of the total wave function leads to

\begin{align}
\langle \Psi \vert \Psi\rangle=1=\sum_n \langle \phi^n \vert \phi^n \rangle,
\end{align}
where the coefficient wave functions $\vert \phi^n \rangle$ are not normalized and form an ensemble that determines $\vert \Psi(t) \rangle$. These coefficients consider the dependence of the total wave function on the degree of freedom of the system. So, they are functions of the system coordinates. Furthermore, we define the projection operators to characterize the ensemble as \cite{Biele,Gaspard}

\begin{subequations}
\begin{align}
&P\equiv \mathbf{1}_{\rm s} \otimes \vert l \rangle \langle l \vert  \\
&Q \equiv \mathbf{1}_{\rm s} \otimes\sum_{n\neq l}\vert n \rangle \langle n \vert,
\end{align}
\end{subequations}
The $l$th coefficient wave function is $P \vert \Psi \rangle=\vert \phi^l \rangle \otimes \vert l \rangle$. In a similar way, $Q \vert\Psi \rangle$ gives the other wave functions in the ensemble. Note that this operators satisfy 

\begin{subequations}
\begin{align}
&P^2=P=P^{\dagger} \\
&Q^2=Q=Q^{\dagger} \\
&P+Q=\mathbf{1}_{\rm T},
\end{align}
\end{subequations}
where T denotes the total wave function.

Now we can use the Feshbach projection-operator method \cite{Nakajima,Zwanzig} to derive a non-Markovian SSE from of Eq. (119). This approach is implemented in the interaction picture as

\begin{align}
i \D \vert \Psi_{\rm I}(t)\rangle=\lambda W(t) \vert \Psi_{\rm I}(t)\rangle \D t,
\end{align}
where

\begin{subequations}
\begin{align}
&\vert \Psi_{\rm I}(t)\rangle=e^{iH_\varepsilon t}e^{iH_{\rm s} t}\vert \Psi (t)\rangle \\
&W(t)=e^{iH_\varepsilon t}e^{iH_{\rm s} t} W e^{-iH_{\rm s} t} e^{-iH_\varepsilon t},
\end{align}
\end{subequations}

In the Feshbach projection method, we decompose the Schrödinger equation into two equations: The first equation gives us information about the time evolution of a member of the ensemble, $P\vert \Psi\rangle$, and the second equation is a differential equation for $Q \vert \Psi\rangle$ and gives information about other coefficient wave functions. We solve the second equation and insert it into the first to obtain a differential equation for $P\vert \Psi\rangle$. 

Based on the mentioned plan, we apply the projection operators in Eq. (123)  to Eq. (125) and after some rearrangement, we obtain \cite{Biele,Gaspard}

\begin{align}
i \D P \vert \Psi_{\rm I}(t)\rangle =\lambda PW(t)P\vert \Psi_{\rm I}(t)\rangle \D t+\lambda PW(t) \Big( U(t)Q\vert \Psi_{\rm I}(0)\rangle  -i\lambda \int_0^t \D t' U(t-t') QW(t') P\vert \Psi_{\rm I}(t')\rangle \Big) \D t,
\end{align}
where $U(t)$ is the time evolution operator and obeys $i \D U(t)=\lambda QW(t) QU(t) \D t$. Note that Eq. (127) shows the exact time evolution of the $l$th coefficient of $\vert \Psi_{\rm I}(t)\rangle$. However, this relation is difficult to solve for the total system. Therefore, we consider a case in which the system is weakly coupled to the environment. Consequently, we perform the second-order perturbation theory in $\lambda$ as

\begin{align}
i \D P\vert \Psi_{\rm I}(t)\rangle &=\lambda PW(t)P \vert \Psi_{\rm I}(t)\rangle \D t+\lambda PW(t) Q \vert \Psi_{\rm I}(0)\rangle \D t \nonumber\\
&-i\lambda^2 PW(t) \int_0^t \D t' \Big(QW(t') Q\vert \Psi_{\rm I}(0)\rangle +QW(t')P \vert \Psi_{\rm I}(t')\rangle\Big) \D t +\mathcal{O}(\lambda^3),
\end{align}

To be more clear, let us consider the interaction potential that has a linear form as $W=\sum_{\alpha} V_{\alpha} \otimes B_{\alpha}$ in which $V_{\alpha} $ and $B_{\alpha}$ denotes the Hermitian operators of the system and environment, respectively. We multiply Eq. (128) with $\langle l \vert$ and assume $\langle l \vert B_{\alpha}(t) \vert l \rangle=0$. So, we have

\begin{align}
i \D  \vert \phi_{\rm I}^l (t)\rangle =\vert f(t) \rangle \D t -i \lambda^2 \sum_{\alpha,\beta} V_{\alpha} (t) \int_0^t \D t' V_{\beta} (t') \langle l \vert B_{\alpha} (t) B_{\beta} (t') \vert l \rangle \vert \phi_{\rm I}^l (t')\rangle \D t,
\end{align}
where $\vert f(t) \rangle$ is the forcing term and shows the effects of all other environmental modes on the $l$th coefficient wave function 

\begin{align}
\vert f(t) \rangle =\lambda \sum_{\alpha,\beta,n (\neq l)} V_{\alpha}(t) \Big [ \langle l \vert B_{\alpha}(t) \vert n \rangle -i \lambda \int_0^t \D t' V_{\beta} (t') \langle l \vert B_{\alpha} (t) B_{\beta}(t') \vert n \rangle \Big] \vert \phi_{\rm I}^n(0) \rangle ,
\end{align}
where the initail condition $\phi_{\rm I}^n(0)$ is appeared in this relation. We assume the system is in a pure state and the environment is in thermal equilibrium at $t=0$. So, the initial total wave function is given by

\begin{align}
\vert \Psi(0) \rangle = \vert \phi(0) \rangle  \otimes \sum_n \sqrt{\frac{e^{-\beta' \epsilon_n}}{Z_\varepsilon}}e^{i \theta_n} \vert n \rangle,
\end{align}
where $\beta'$ is the inverse of the temperature and $Z_\varepsilon =\tr _\varepsilon e^{-\beta' H_\varepsilon}$. In addition, $\lbrace \theta_n\rbrace$ are independent random phases uniformly distributed over $[0,2\pi]$. Here, contrary to the master equation approach, the random phase factors need to be taken into account in the initial conditions. So, one can obtain

\begin{align}
\vert \phi^n(0)\rangle= \langle n \vert \Psi(0)\rangle=\vert \phi^l(0)\rangle e^{-\frac{\beta'}{2} (\epsilon_n-\epsilon_l)}e^{i(\theta_n-\theta_l)},
\end{align}
Consequently, after some manipulations, we obtain 

\begin{align}
\vert f(t)\rangle= \lambda \sum_{\alpha} \gamma_{\alpha}^l (t) V_{\alpha} (t)  \vert \phi_{\rm I}^l (t) \rangle,
\end{align}
where the stochastic noise $\gamma_{\alpha}^l(t)$ is obtained after a thermal average as

\begin{align}
\gamma_{\alpha}(t)=\frac{1}{\sqrt{Z_\varepsilon}} \sum_{l,n \neq l} \langle l \vert B_{\alpha} (t) \vert n \rangle e^{-\frac{\beta'}{2} \epsilon_n}e^{i(\theta_n-\theta_l)} ,
\end{align}
Moreover, we assume 

\begin{align}
\langle l \vert B_{\alpha} (t) B_{\beta} (t') \vert l \rangle \equiv C_{\alpha,\beta} (t-t'),
\end{align}
where $C_{\alpha,\beta} (t-t')$ is the environmental correlation function. We consider the environment to be large enough and therefore $\gamma_{\alpha}(t)$ including a sum of oscillating terms leads to a random Gaussian behavior. Hence, the noise is determined by the mean value and variance. 

Considering all mentioned information, Eq. (129) can be written as

\begin{align}
i \D  \vert \phi (t)\rangle =H_{\rm s}  \vert \phi (t)\rangle \D t+\lambda \sum_{\alpha} \gamma_{\alpha} (t) V_{\alpha}  \vert \phi (t)\rangle \D t -i \lambda ^2 \sum_{\alpha,\beta} V_{\alpha} \int_0^t \D \tau e^{-i H_{\rm s} \tau} V_{\beta}  C_{\alpha,\beta} (\tau)  \vert \phi (t-\tau)\rangle \D t,
\end{align}
where is a non-Markovian SSE \cite{Biele,Gaspard}.

For simplicity, one can use the $\delta$-correlated environment approximation as $C_{\alpha,\beta} (t-t') \approx  D_{\alpha,\beta} \delta (t-t')$ to neglect the non-Markovian behavior of the system. Consequently, we obtain the Markovian SSE in Ito differential form \cite{Karatzas,Protter} as

\begin{align}
\D \vert \phi (t)\rangle =\Big\lbrace \Big[ -i H_{\rm s}-\frac{1}{2} \sum_{\alpha} S_{\alpha}^{\dagger} S_{\alpha}\Big] \D t + \sum_{\alpha} S_{\alpha} \D W_{\alpha}\Big\rbrace \vert \phi (t)\rangle ,
\end{align}
where the new environmental operators are $S_{\alpha}= \lambda \sqrt{d_{\alpha}} \sum_{\beta} U_{\alpha,\beta} V_{\beta}$ where $U_{\alpha,\beta}$ diagonalizes $D_{\alpha,\beta}$ with eigenvalues $d_\alpha$ \cite{Biele}.

In summary, to obtain SSE, we have built an ensemble of states of the system each of which evolves independently and can be identified by the dynamics of the environment. Moreover,  we perform an average over this ensemble to calculate any physical quantity. Note that many approaches have been developed to deriving SSE in open quantum systems. For more details, please see \cite{Breuer,Wiseman1,Wiseman2,Wiseman3,Carmichael,Diosi2}.

There are some standard techniques to solve SSE numerically for the realization of the noise such as the Euler scheme, the Heun scheme and the fourth-order Runge-Kutta scheme, etc \cite{Breuer,Kloeden,Higham,Burrage,Billah}. The details of each of these methods are beyond the scope of this paper.

\subsubsection{Stochastic master equation (SME)}

The stochastic master equation (SME) is a general and useful approach to describe the interaction between a quantum system such as a qubit and a fluctuating environment such as quantum noise \cite{Wiseman4,Steck}. There are many methods to deriving SME based on the features of the problem in question \cite{Wiseman4,Lacroix, Steck,Rouchen,Braginsky,Fleming}, but here we derive it according to the formalism of the previous section.

We have discussed the framework of SSE to analyze the dynamics of an open quantum system based on the state of the system $ \vert \phi (t)\rangle$. However, for simplicity and practical reasons it is better, that we formulate the stochastic approach using the statistical operator or the reduced density matrix of the system which is defined as 

\begin{align}
\rho\equiv \overline{\vert \phi \rangle \langle \phi \vert},
\end{align}
where $\overline{...}$ denotes the mean value. Moreover, we need to use the Ito chain rules \cite{Gardiner} which are defined as

\begin{align}
\D (\vert \phi \rangle \langle \psi \vert)=(\D \vert \phi \rangle) \langle \psi \vert  +\vert \phi \rangle (\D \langle \psi \vert)+(\D \vert \phi \rangle) (\D \langle \psi \vert),
\end{align}

So, we have \cite{Biele}

\begin{subequations}
\begin{align}
&\overline{\D W_{\alpha} \D W_{\beta}^*}=\delta_{\alpha,\beta} \D t \\
&\overline{\D W_{\alpha} \D W_{\beta}}=0 \\
&\overline{\D W_{\alpha} \D t}=0,
\end{align}
\end{subequations}
where $\vert \phi \rangle$ and $\vert \psi \rangle$ are two states evolving based on Eq. (137).

To derive the equation of motion for $\rho$ corresponding to the Markovian SSE in Eq. (137), we calculate $\D \rho$ as

\begin{align}
\D \rho=\D \overline{\vert \phi \rangle \langle \phi \vert} =\overline{(\D \vert \phi \rangle) \langle \phi \vert + \vert \phi \rangle  (\D \langle \phi \vert)+ (\D \vert \phi \rangle) (\D \langle \phi \vert)            },
\end{align}
In Markovian regime, using Eqs. (139)-(141) one can show \cite{Biele}

\begin{align}
\D \rho =- i \big [H_{\rm s}, \rho \big]\D t-\frac{1}{2} \sum_{\alpha} \Big \lbrace S_{\alpha}^{\dagger} S_{\alpha} \rho+\rho S_{\alpha}^{\dagger} S_{\alpha}  -2 S_{\alpha} \rho S_{\alpha}^{\dagger} \Big\rbrace \D t+\mathcal{O}(\D t^2),
\end{align}
where is the most general type of a Markovian master equation called the Lindblad master equation. Therefore, we conclude the Markovian SSE in Eq. (137) explains the dynamics of an open quantum system on average which finally obtains the Lindblad equation. Note that if we consider a stochastic Hamiltonian for the system, we will have to deal with a set of Hamiltonians and the equation of motion will be different from the Lindblad master equation.

To derive a non-Markovian SME corresponds to Eq. (136), we perform a perturbative expansion in $\lambda$ up to forth order \cite{Gaspard}. Moreover, we assume the Hamiltonian is non-stochastic and consider the limit $t\rightarrow \infty$ in the history term. After some calculations, the non-Markovian SME is obtained as

\begin{align}
\frac{\D \rho(t)}{dt}=-i \big[ H_{\rm s},\rho(t) \big]+\sum_{\alpha} \Big( K_{\alpha} \rho (t) V_{\alpha}+V_{\alpha} \rho(t) K_{\alpha}^{\dagger}  -V_{\alpha} K_{\alpha}\rho(t)-\rho(t)K_{\alpha}^{\dagger}V_{\alpha} \Big),
\end{align}
where

\begin{align}
K_{\alpha}=\lambda^2 \sum_{\beta} \int_0^{\infty} \D \tau C_{\alpha,\beta} (\tau) e^{-iH_{\rm s}\tau} V_{\beta} e^{iH_{\rm s}\tau}.
\end{align}
where is a time-local equation called the Redfield master equation \cite{Redfield}. This equation can be applied in a situation in which the dynamics of the environment is much faster than the system dynamics.

In conclusion, it seems that SSEs can play an important role in obtaining and solving the corresponding SME. In the next subsection, we analyze the dynamics of superconducting qubits based on the stochastic approach.

\subsubsection{Stochastic approach in superconducting qubits}

In recent years, stochastic approaches have attracted much attention in the analysis of superconducting qubits technology such as CQED and the relationship between quantum fluctuations and measurement outcomes. In this context, the measurement of a superconducting qubit is modeled using stochastic master equations 
\cite{Yu2,Tornberg,Ibarcq,Cernotik,Roch}. Moreover, in some studies, stochastic approaches have been used to analyze the dissipative behavior of superconducting qubits interacting with environmental noises \cite{Ficheux,Gustavsson,Gambetta,Schwartz}.

Here, we describe the stochastic approaches for the qubit-cavity coupled system and the continuous measurement that fundamentally disturbs the qubit based on \cite{Jacobs2,Naghiloo2}.

\subsubsection*{i. SSE}

Let us analyze the qubit evolution under the measurement operator $\Omega_{\tilde{V}}$ which can be defined as \cite{Naghiloo2}

\begin{align}
\Omega_{\tilde{V}} \simeq e^{-k \Delta t (\tilde{V}-\sigma_z)^2},
\end{align}
where $\sigma_z$ and $k$ denote the spin Pauli of the qubit and the cavity linewidth in the qubit-cavity system, respectively. Moreover, $\tilde{V}$ is the measurement signal that we assume has a Gaussian distribution centered on $\sigma_z$ as

\begin{align}
\tilde{V}= \langle \sigma_z \rangle+\frac{\D \mathcal{W}}{\sqrt{4k \Delta t}},
\end{align}
where $\D \mathcal{W}$ is a Wiener increment which is a zero-mean Gaussian random variable. The Wiener process $\mathcal{W}(t)$ is a type of real-valued continuous-time stochastic process used to study the characteristics of one-dimensional Brownian motion. It can be considered an ideal random walk with independent and small steps \cite{Jacobs2}.

We consider $\vert \psi(t)\rangle$ as the normalized pure state of the qubit at time $t$. At time $t+\Delta t$, the qubit state evolves as

\begin{align}
\vert \psi(t+\Delta t)\rangle &=\Omega_{\tilde{V}} \vert \psi(t)\rangle \nonumber\\
&\propto e^{-k \Delta t (\tilde{V}-\sigma_z)^2}  \vert \psi(t)\rangle  \nonumber\\
&\propto e^{-k \Delta t (\sigma_z ^2-2 \tilde{V} \sigma_z )} \vert \psi(t)\rangle,
\end{align}
where we have eliminated the constant term $\tilde{V} ^2 $, because we want to renormalize $\vert \psi(t+\Delta t)\rangle $ eventually. Substituting Eq. (146) into Eq. (147), and replacing $\Delta t\rightarrow \D t$, we obtain

\begin{align}
\vert \psi(t+\D t)\rangle =\Big(1-k\D t \sigma_z^2 +2k \D t \sigma_z \langle \sigma_z \rangle  +\sqrt{k} \sigma_z \D \mathcal{W}+\frac{k}{2} \sigma_z^2 (\D \mathcal{W})^2\Big) \vert \psi(t)\rangle ,
\end{align}
up to the first order in $\D t$. Considering the stochastic Ito rule, we use $(\D \mathcal{W})^2=\D t$ \cite{Jacobs2}. So, we have

\begin{align}
\vert \psi(t+\D t)\rangle &=\Big(1-\frac{k}{2} \sigma_z [\sigma_z - 4 \langle \sigma_z \rangle ] \D t+\sqrt{k} \sigma_z  \D \mathcal{W}\Big) \vert \psi(t)\rangle ,
\end{align}

To obtain the normalized state of the qubit, one can show \cite{Naghiloo2}

\begin{align}
\langle \psi(t+\D t)\rangle  \vert \psi(t+\D t)\rangle = 1+ 4 k \langle \sigma_z \rangle ^2 \D t +\sqrt{4k} \langle \sigma_z \rangle \D \mathcal{W}+ \mathcal{O}(t)^{3/2},
\end{align}
up to the second order in $\D \mathcal{W}$. We use the binomial expansion and show

\begin{align}
\Big[ \langle \psi(t+\D t)\rangle  \vert \psi(t+\D t)\rangle \Big]^{-\frac{1}{2}}&=1-\frac{k}{2} \langle \sigma_z \rangle ^2 \D t -\sqrt{k} \langle \sigma_z \rangle \D \mathcal{W} +\mathcal{O}(t)^{3/2},
\end{align}
Consequently, we multiply Eq. (151) by Eq. (149) and obtain the normalized SSE as

\begin{align}
\D \vert \psi(t) \rangle =\Big(\frac{-k}{2} [ \sigma_z-\langle \sigma_z \rangle ]^2 \D t+\sqrt{k} [ \sigma_z-\langle \sigma_z \rangle ]  \D \mathcal{W} \Big) \vert \psi(t) \rangle,
\end{align}
where  $\D \vert \psi(t) \rangle=\vert \psi(t+\D t) \rangle-\vert \psi(t) \rangle$.

For a specific measurement, we can obtain $ \D \mathcal{W}$  from Eq. (146) and then integrate SSE in Eq. (152) to calculate the evolution of the pure state of qubit under the measurement.

\subsubsection*{ii. SME}

The SSE in Eq. (152) can be used for the evolution of the pure state of the qubit. To describe the mixed state evolution, we generalize SSE in terms of the density matrix. In this regard, we use

\begin{subequations}
\begin{align}
&\rho=\vert \psi \rangle \langle \psi \vert  \\
& \D \rho= \D \vert \psi \rangle \langle \psi \vert +\vert \psi \rangle \D \langle \psi \vert +\D \vert \psi \rangle \D \langle \psi \vert,
\end{align}
\end{subequations}
By substituting $\D \vert \psi \rangle$ from Eq. (152), we can obtain the SME as  \cite{Naghiloo2}

\begin{align}
\D \rho= -\frac{k}{2} [\sigma_z,[\sigma_z,\rho]] \D t +\sqrt{k} (\sigma_z \rho+\rho \sigma_z-2 \langle \sigma_z \rangle \rho )\D \mathcal{W}.
\end{align}
where we have a double commutator. Note that the SME can be used for any $\rho$, either pure or mixed.

In general, SME can be divided into three parts: the first is the Heisenberg equation of motion of the qubit. The second is the dissipative evolution of qubit in the Lindblad form. The last part is due to the stochastic perturbation of the qubit by the measurement process \cite{Vool}.

\begin{figure*}[h]
\centering
\includegraphics[scale=0.25]{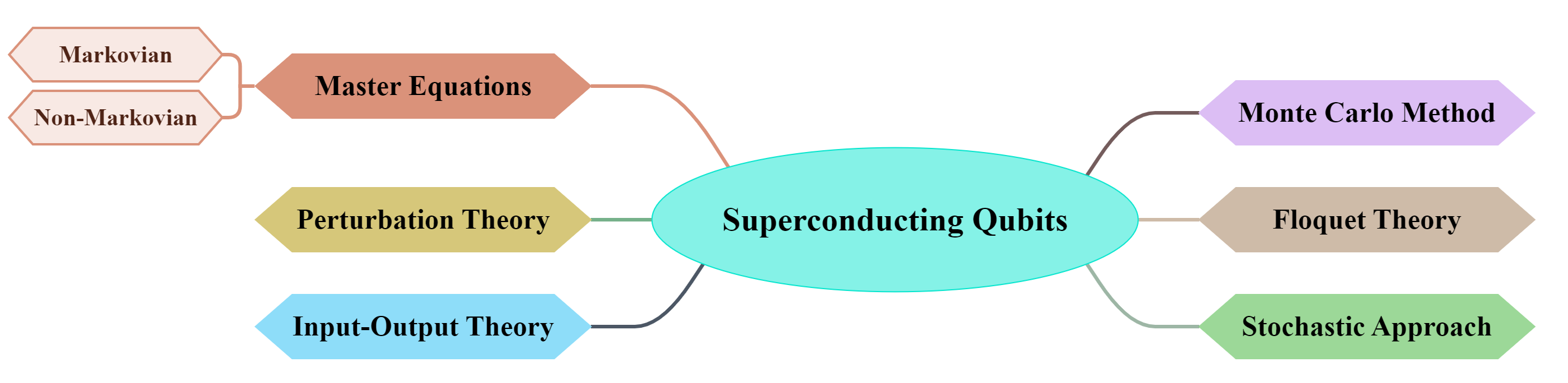}
\caption{Open quantum system approaches to analyze the dynamics of superconducting qubits interacting with the environment which are discussed in this work.}
\end{figure*}

\section{Summary}

Like all types of qubits, superconducting qubits have their challenges to be applicable in practical quantum computing, whether in noisy intermediate-scale quantum computing (NISQ) or fault-tolerant quantum computers (FTQC). One of the most important challenges is the decoherence effects of the environmental noises which causes the reduction of qubit coherence time as well as gate fidelities.
Therefore, decoherence poses a major obstacle to reaching practical quantum computers. In addition to designing robust qubits against environmental noises, developing different methods to investigate the dynamics of a superconducting qubit as an open quantum system is vital. 

In this study, we have discussed various open quantum system approaches to model superconducting qubit dynamics under environmental noise that are useful for qubit design, control, and readout. These methods are given in FIG. 8. Each presented method is suited for specific noise types and qubit-environment interactions, providing insights into qubit dynamics. We aimed to explain in detail how these methods specifically apply to superconducting qubits. These techniques can help to shed light on decoherence mechanisms and qubit dynamics in CQED and related architectures. They remain indispensable for quantum noise engineering and mitigation efforts for robust quantum information processing.  The choice of approach often depends on the specific characteristics of the qubit-environment interaction being studied: phenomenological Lindblad equations are useful for general descriptions of decay and dephasing, Bloch-Redfield connects these rates to microscopic noise spectra, non-Markovian methods are crucial for memory effects, Floquet theory for periodically driven systems, input-output theory for systems coupled to transmission lines and measurement apparatus, and stochastic methods for simulating individual quantum trajectories or specific noise models.

While this review does not detail specific experimental coherence enhancements resulting from each theoretical method, it underscores that a deep understanding of qubit-environment interactions, facilitated by the discussed open quantum system approaches, is a prerequisite for designing robust qubits and developing effective error mitigation and correction strategies that lead to such improvements. Developing different methods to investigate the dynamics of a superconducting qubit as an open quantum system is vital for informing these advancements. Considering the wide range of theoretical and experimental studies in this field, we cannot claim to have considered all possible methods, but we believe that this work can provide a comprehensive overview of important methods that can be considered by researchers.

\subparagraph{Acknowledgment}

The author would like to thank Mojtaba Tabatabaei for his useful comments, Mohsen Akabri for funding, and Mohammad Mirsadeghi for helping in preparing some figures.

%\subparagraph{Authors Contribution}

%\subparagraph{Conflict of Interest}

%The authors have no conflicts to disclose.

\subparagraph{Data Availability}

The data that support the findings of this study are available on request from the corresponding author.

\end{document}